\documentclass[aps,prd,twocolumn,twoside,floatfix,superscriptaddress,preprintnumbers]{revtex4}

\usepackage{graphicx}
\usepackage{amsmath}

\usepackage{enumerate}

\newcommand{\bea}{\begin{eqnarray}}
\newcommand{\beq}{\begin{equation}}
\newcommand{\eea}{\end{eqnarray}}
\newcommand{\eeq}{\end{equation}}

\newcommand{\lsim}{\stackrel{<}{_\sim}}

\newcommand{\be}{\begin{equation}}
\newcommand{\ee}{\end{equation}}

\newcommand{\ba}{\begin{array}}
\newcommand{\ea}{\end{array}}

\def\R1{\varepsilon_1}
\def\E8{\varepsilon_8}

\newcommand{\bd}{\begin{displaymath}}
\newcommand{\ed}{\end{displaymath}}

\newcommand{\bi}{\begin{itemize}}
\newcommand{\ei}{\end{itemize}}

\newcommand{\sle}{{\tilde\ell}}
\newcommand{\nt}{{\tilde\chi}^0}
\newcommand{\ch}{{\tilde\chi}}

\newlength{\myem}
\newcommand{\sep}[1]{#1}
\newcounter{mysubequation}[equation]
\renewcommand{\themysubequation}{\alph{mysubequation}}
\newcommand{\mytag}{\stepcounter{mysubequation}%
\tag{\theequation\protect\sep{\themysubequation}}}
\newcommand{\globallabel}[1]{\refstepcounter{equation}\label{#1}}

\begin{document}
\preprint{TUM-HEP-742/09}
\preprint{MPP-2009-208}

\title{Slepton mass-splittings as a signal of LFV at the LHC}

\author{Andrzej~J.~Buras}
\affiliation{Physik-Department, Technische Universit\"at M\"unchen,
D-85748 Garching, Germany}
\affiliation{TUM Institute for Advanced Study, Technische Universit\"at M\"unchen,
\\Arcisstr.~21, D-80333 M\"unchen, Germany}
\author{Lorenzo~Calibbi}
\affiliation{Max-Planck-Institut f\"ur Physik (Werner-Heisenberg-Institut), D-80805 M\"unchen, Germany}
\author{Paride~Paradisi}
\affiliation{Physik-Department, Technische Universit\"at M\"unchen,
D-85748 Garching, Germany}

\begin{abstract}

Precise measurements of slepton mass-splittings might represent a powerful tool to probe 
supersymmetric (SUSY) lepton flavour violation (LFV) at the LHC. We point out that 
mass-splittings of the first two generations of sleptons are especially sensitive to LFV 
effects involving $\tau-\mu$ transitions. If these mass-splittings are LFV induced, 
high-energy LFV processes like the neutralino decay ${\nt}_2\to\nt_1\tau^{\pm}\mu^{\mp}$ 
as well as low-energy LFV processes like $\tau\to\mu\gamma$ are unavoidable. We show that 
precise slepton mass-splitting measurements and LFV processes both at the high- and low-energy 
scales are highly complementary in the attempt to (partially) reconstruct the flavour sector 
of the SUSY model at work. The present study represents another proof of the synergy and 
interplay existing between the LHC, i.e. the {\em high-energy frontier}, and high-precision 
low-energy experiments, i.e. the {\em high-intensity frontier}.
\end{abstract}

\maketitle

\section{Introduction}\label{sec:intro}

The most important achievement we expect to reach at the beginning of the LHC era is the
understanding of the underlying mechanism accounting for the electroweak symmetry breaking,
in particular, whether the Higgs mechanism is realized in nature or not.
Moreover, the LHC is also expected to shed light on the hierarchy problem, since a natural
solution of it calls for a TeV scale New Physics (NP).

On the other hand, low-energy flavour physics observables provide the most powerful tool 
to unveil the symmetry properties of the NP theory that will emerge at the LHC, if any.
In fact, high-precision measurements at the LHC are made typically challenging by the huge 
background and by irreducible hadronic uncertainties.

A remarkable exception, arising in SUSY theories, is given by the possibility to access
information about SUSY masses relying on some kinematical observables, as the kinematic
end-point of the invariant mass distribution of the leptons coming from
neutralino-slepton-neutralino cascade decays \cite{edge},
${\nt}_2 \to \sle^{\pm}\ell^{\mp} \to \nt_1 \ell^{\pm} \ell^{\mp}$. If the slepton in
the decay chain is real, the di-lepton invariant mass spectrum has a prominent kinematic
edge~\cite{edge} that may be measured at the LHC experiments with very high precision
(up to 0.1 \%)~\cite{edge,Armstrong:1994it}, allowing, in combination with other kinematical observables, 
to reconstruct the masses of the particles involved in the chain, in particular
the slepton masses~\cite{edge}.

In this work, we point out that precise measurements of slepton mass-splittings might represent 
a powerful tool to probe LFV at the LHC.

In particular, we consider minimal gravity mediated SUSY breaking scenarios (mSUGRA), where
the first two slepton generations are predicted to be highly degenerate, typically below the
percent level.
Hence, any experimental evidence for a sizable mass splitting between selectrons and smuons 
($\Delta m_{\sle}/m_{\sle}$), say above the percent level, would signal either a different 
mechanism for SUSY breaking or non minimal realizations of SUGRA breaking models.
In the latter case, the presence of LFV interactions might be at the origin of such a significant
mass splitting $\Delta m_{\sle}/m_{\sle}$.

In this context, we point out that sizable values for $\Delta m_{\sle}/m_{\sle}$ can be generated only through flavour mixings between the {\it second} and {\it third} slepton families, as it
might naturally arise from the large mixing angle observed in atmospheric neutrino oscillation experiments. In contrast, flavour mixings between the second and first slepton families are
tightly constrained by the non observation of $\mu\to e\gamma$, hence, they cannot induce testable values for $\Delta m_{\sle}/m_{\sle}$, unless very special and fine-tuned conditions are fulfilled \cite{Hisano:2002iy}.

If these mass-splittings are LFV induced, high-energy LFV processes like the neutralino decay
${\nt}_2\to\nt_1\tau^{\pm}\mu^{\mp}$ \cite{arkani,hinchliffe,ellis} 
as well as low-energy LFV processes like $\tau\to\mu\gamma$
are unavoidable. We show that precise slepton mass-splitting measurements and LFV processes
(both at the high- and low-energy scales) are highly complementary in the attempt to (partially)
reconstruct the flavour sector of the SUSY model at work.

Our paper is organized as follows. In Section~\ref{Sec:2} we summarize those slepton mass measurements
at the LHC that are relevant for our paper. Subsequently in Section~\ref{Sec:3} we discuss various
aspects of LFV at the LHC. In Section~\ref{Sec:4} we present the numerical analysis of LFV and in
Sections~\ref{Sec:5} and \ref{sec:6} 
estimates of the cross-sections, expected number of events at the LHC and backgrounds 
for the relevant processes discussed in our paper are made. 
In Section~\ref{Sec:7} we summarize our main findings.

\section{Slepton mass measurements at the LHC}\label{Sec:2}

Sleptons can be produced at the LHC in two possible ways: either directly in quark collisions, through Drell-Yan s-channel $Z^0/\gamma$ exchange, or indirectly from cascade decays of squarks
and gluinos through neutralinos. In the case of direct production, the detection of sleptons is
made challenging by relatively low cross-section and the large SM background and it should be
feasible only for slepton masses up to 200-300 GeV \cite{Drell-Yan}. Moreover, the indetermination
of the center of mass energy of the parent quarks makes really difficult to extract information
about the slepton masses. On the other hand, sleptons can be copiously produced in cascade decays
of squarks through neutralinos, if the processes in the chain are kinematically allowed. Within several SUSY models, such as the CMSSM, where the second-lightest neutralino is mostly Wino-like,
one of the most effective of such chains is
\begin{equation}
{\tilde q}_L \to q_L\, \nt_2 \to q_L \,\sle^\pm \ell^\mp \,,
\label{cascade1}
\end{equation}
with typically ${\rm BR}({\tilde q}_L \to q_L\, \nt_2)\simeq 1/3$. The slepton will typically decay 
into a lepton and the LSP. Besides the possibly large amount of sleptons produced in this way (clearly if 
$m_\sle< m_{\nt_2}$), the main advantage with respect to the direct Drell-Yan production is given 
by the possibility to access information about SUSY masses relying on some kinematical observables, 
which may be measured very precisely at the LHC.

An example of such observables, which is particularly important in the case of sleptons, is given by
the kinematic end-point of the invariant mass distribution of the leptons coming from 
neutralino-slepton-neutralino cascade decays \cite{edge},
\begin{equation}
{\nt}_2 \to \sle^{\pm}\ell^{\mp} \to \nt_1 \ell^{\pm} \ell^{\mp}\,.
\label{cascade}
\end{equation}
If the slepton in the decay chain is real, the di-lepton invariant mass spectrum has a 
prominent kinematic edge \cite{edge} at
\begin{equation}
m_{ll}^2 =  \frac{(m_{\nt_2}^2 - m_{\tilde{\ell}}^2)
(m_{\tilde{\ell}}^2 - m_{\nt_1}^2)}{m_{\tilde{\ell}}^2}.
\label{eq:edge}
\end{equation}
Such an edge may be measured at the LHC experiments with very high precision (up to 0.1 \%)~\cite{edge,Armstrong:1994it}, allowing, in combination with other kinematic observables,
to reconstruct the masses of the particles involved in the chain, in particular the slepton
masses \cite{edge}.

Interestingly, in the region of the parameter space where decays of neutralino in real sleptons
are kinematically allowed, $\rm{BR}({\nt}_2\to\nt_1\ell^{\pm}\ell^{\mp})$ is clearly enhanced 
with respect to the case of virtual intermediate sleptons. Noteworthy enough, in the CMSSM, such
region of the parameter space is close to the region where $m_{\tilde\tau_{1}}\simeq m_{\nt_1}$
(where ${\tilde\tau_{1}}$ is the lightest stau) and the WMAP dark-matter (DM) constraints are naturally satisfied by an efficient ${\tilde\tau_{1}}$-LSP coannihilation \cite{ellis}.

For our purposes, it is important to understand whether the measurement of the kinematic
edge of Eq.~(\ref{eq:edge}) in the case of $\mu-\mu$ and $e-e$ mass distributions can be
used to resolve a (small) mass difference between the corresponding sleptons ($\tilde \mu$
and $\tilde e$, in our case).

The fractional shift in the invariant mass edge in terms of the slepton mass splitting 
$\Delta m_{\tilde{\ell}}/m_{\tilde{\ell}}$ is given by
\beq
\frac{\Delta m_{ll}}{m_{ll}} = \frac{\Delta m_{\tilde{\ell}}}{m_{\tilde{\ell}}}
\left( \frac{m_{\nt_1}^2 m_{\nt_2}^2 - m_{\tilde{\ell}}^4}{(m_{\nt_2}^2 - m_{\tilde{\ell}}^2 )
(m_{\tilde{\ell}}^2 - m_{\nt_1}^2)} \right)\,.
\label{edgesplitting}
\eeq
where the factor multiplying $\Delta m_{\tilde{\ell}}/m_{\tilde{\ell}}$ 
can provide an enhancement of the edge splitting~\cite{Allanach:2008ib}.

The above equation deserves some comments: i) when $m_{\tilde{\ell}}=
\sqrt{m_{\nt_1}m_{\nt_2}}$ the shift in the edge vanishes to leading order in
$\Delta m_{\sle}/m_{\sle}$, ii) large splittings of the di-lepton edges can be
achieved for relatively small splittings of the selectron and smuon masses and
iii) the enhancement factor is larger for more degenerate masses of sparticles
in the chain and can easily be $\mathcal{O}(10)$ depending upon the value of $m_{\nt_2}/m_{\nt_1}$~\cite{Allanach:2008ib}. 
The enhancement diverges as the slepton mass approaches either neutralino masses.
In the latter case, such an enhancement is not effective because the leptons coming
from such chain tend to be soft and thus hard to detect and identify experimentally.

\subsection{Lepton flavour conserving case}

Within the CMSSM, the first two slepton generations are degenerate to a very large extent.
In particular, in the absence of flavour mixing angles, the slepton masses $m_{\sle_{1,2}}$ are
given by
\beq
m^{2}_{\sle_{1,2}} \!=\! \frac{(m^{2}_{\sle_{L}} \!+\! m^{2}_{\sle_{R}})}{2}\mp
\frac{\sqrt{(m^{2}_{\sle_{L}} \!-\! m^{2}_{\sle_{R}})^2 + 4(\Delta^{\sle\sle}_{RL})^2}}{2}
\label{slepton_masses}
\eeq
where $m_{\sle_{L(R)}}$ is the left-left (right-right) entry in the slepton mass matrix and
$\Delta^{\sle\sle}_{RL}=m_{\ell}(A_{\ell}-\mu\tan\beta)$ is the left-right mixing term.

At leading order, $m^{2}_{\sle_L}\approx m_{0}^{2}(1-|c|y_{\ell}^{2})+0.5 M^{2}_{1/2}$ and $m^{2}_{\sle_R}\approx m_{0}^{2}(1-2 |c|y_{\ell}^{2}) + 0.15 M^{2}_{1/2}$ where $m_0$ and
$M_{1/2}$ are the universal soft sfermion and gaugino masses while $|c|$ stems from RGE
effects driven by the Yukawa interactions and reads $|c|\approx(3+a_0^2)\ln(M_X/M_Z)/(4\pi)^{2}$
(where $M_X$ is the high scale where the universality is imposed, i.e. either the Planck 
or the GUT scale and $a_0\equiv A_0/m_0$ is the universal trilinear coupling).

Since in the CMSSM $M_{\nt_2}\simeq 0.8 M_{1/2}$, the kinematic condition $m_{\nt_2}>m_{\sle}$
is satisfied when $m_0\lsim 0.35 M_{1/2}$ in case of LH sleptons, $m_0\lsim 0.7 M_{1/2}$ in 
case of RH sleptons. In this regime, it is straightforward to check that
$(m^{2}_{\sle_{L}}\!-\!m^{2}_{\sle_{R}})\gg \Delta^{\sle\sle}_{RL}$
and Eq.~(\ref{slepton_masses}) leads to
\beq
\frac{\Delta m_{\tilde{\ell}}}{m_{\tilde{\ell}}}
\simeq
\frac{m_{\tilde{e}_{R}} - m_{\tilde{\mu}_{R}}}{m_{\tilde{\ell}}}+
\frac{(\Delta^{\tilde{\mu}\tilde{\mu}}_{RL})^2}{m_{\tilde{\ell}}^{2}
(m^{2}_{\tilde{\mu}_{L}}\!-\!m^{2}_{\tilde{\mu}_{R}})}
\label{slepton_masses_approx}
\eeq
where we have defined $m_{\tilde{\ell}}=(m_{\tilde{e}_{R}}+m_{\tilde{\mu}_{R}})/2$. Hence,
even for very large $\tan\beta$ values, the mass splitting between selectrons and smuons,
$\Delta m_{\tilde{\ell}}/m_{\tilde{\ell}}$, can reach at most the per mill level. Therefore,
it is commonly assumed that $m_{ee}$ and $m_{\mu\mu}$ edges occur at identical values in the
CMSSM.

In principle, the enhancement factor for the edge splitting discussed in the above
section could still bring the edge splittings at the percent level.

\begin{figure}[t]
\includegraphics[scale=0.55, angle=-90]{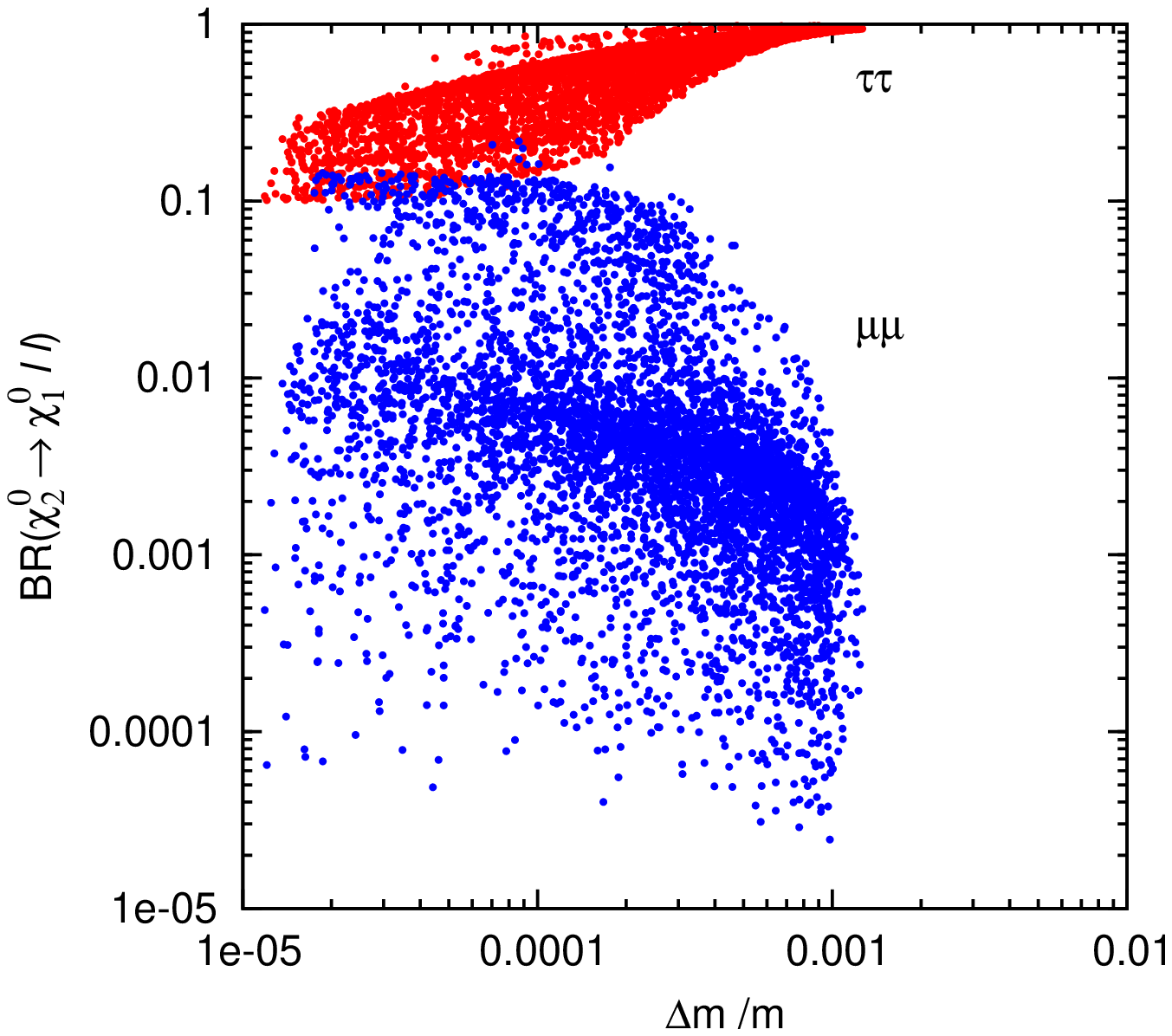}
\includegraphics[scale=0.55, angle=-90]{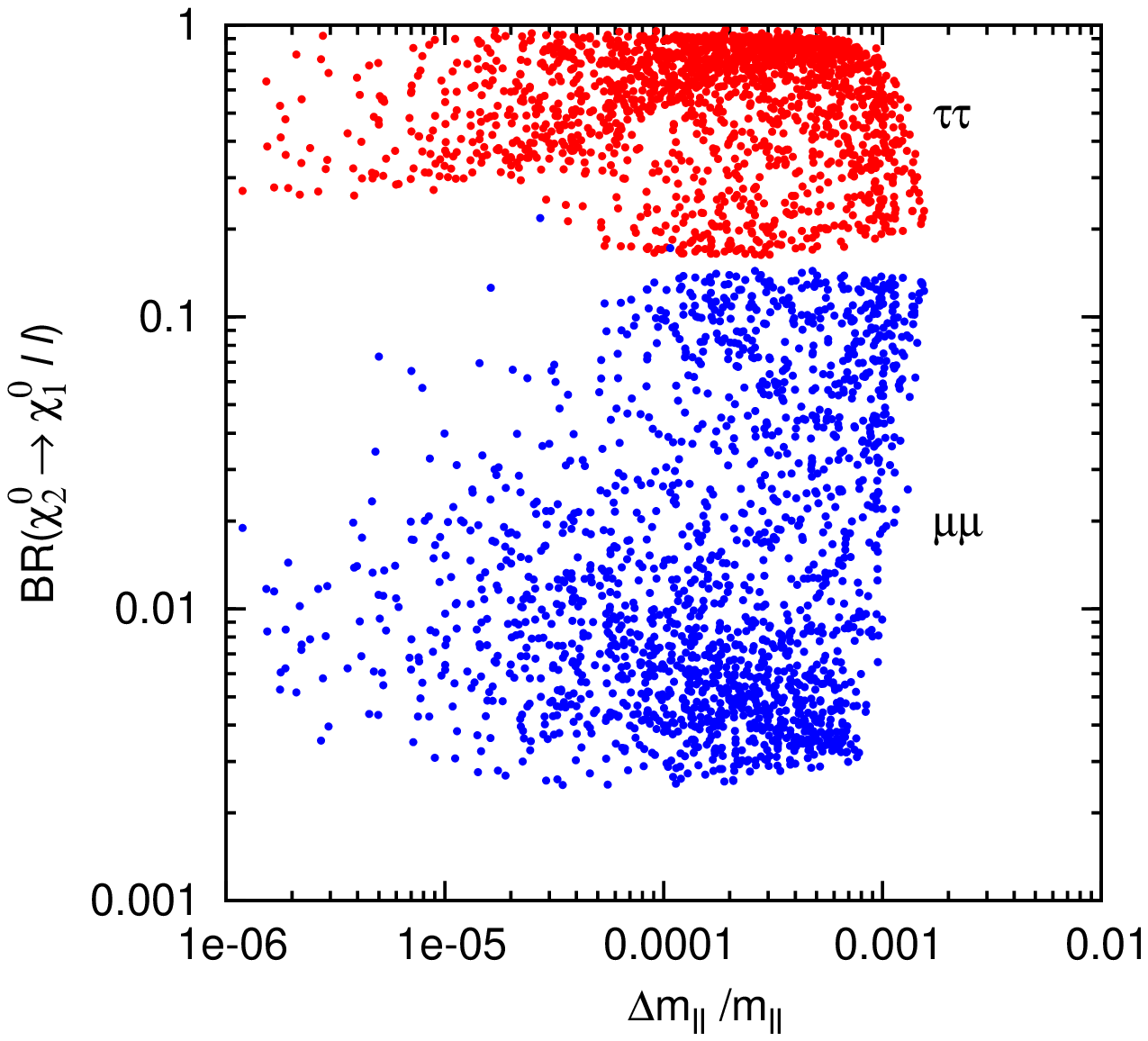}
\caption{${\rm BR}(\nt_{2}\to\nt_{1}\ell\ell)$ (with $\ell=\tau,\mu$) vs. the slepton
mass splitting $\Delta m_{\tilde{\ell}}/m_{\tilde{\ell}}$ (top) and the edge 
splitting $\Delta m_{ll}/m_{ll}$ (bottom).}
\label{mass-edge_splitting}
\end{figure}

In practice, since in the CMSSM $\tilde{\tau}_R$ is driven light at large $\tan\beta$
and, thus, it dominates the $\nt_2$ decay modes, it turns out that ${\rm BR}(\nt_{2}\to\tilde{\tau}_{1}\tau)\sim 1$ 
while ${\rm BR}(\nt_{2}\to\tilde{\ell}_R\ell)\ll 1$
(with $\tilde{\ell}_R={\tilde e}_R,{\tilde\mu}_R$), hence, the edge splitting turns out
to be hardly measurable~\cite{Allanach:2008ib}.
The above situation is well illustrated by Fig.~\ref{mass-edge_splitting} where, on the top,
we show ${\rm BR}(\nt_{2}\to\nt_{1}\ell\ell)$ (with $\ell=\tau,\mu$) vs. the slepton mass
splitting $\Delta m_{\tilde{\ell}}/m_{\tilde{\ell}}$ while on the bottom we show
${\rm BR}(\nt_{2}\to\nt_{1}\ell\ell)$ vs. the edge splitting $\Delta m_{ll}/m_{ll}$
(in this last case we impose the additional constraint of not too soft leptons, namely that
$m_{\nt_{2}}-m_{\tilde{\ell}}\geq 10~\rm{GeV}$ and $m_{\tilde{\ell}}-m_{\nt_{1}}\geq 10~\rm{GeV}$).
The plots of Fig.~\ref{mass-edge_splitting} have been obtained by means of a scan over 
the following ranges of the SUSY parameters: $m_0,M_{1/2}\leq$~1 TeV, $-3\leq a_0\leq +3$ 
and $\tan\beta\leq 50$.
We see that for increasing values of $\Delta m_{\tilde{\ell}}/m_{\tilde{\ell}}$,
${\rm BR}(\nt_{2}\to\nt_{1}\tau\tau)$ increases as well while ${\rm BR}(\nt_{2}\to\nt_{1}\mu\mu)$
is suppressed. In any case, $\Delta m_{\tilde{\ell}}/m_{\tilde{\ell}}$ and $\Delta m_{ll}/m_{ll}$
are very small and they never exceed the per mill level.

We conclude that, any experimental evidence for a sizable mass splitting between $\tilde e$
and $\tilde\mu$ would be a clear signal of a different mechanism for SUSY breaking than the
CMSSM or non minimal realizations of SUGRA breaking models.

\subsection{Lepton flavour violating case}

As discussed before, within minimal SUGRA models, the first two slepton generations are
highly degenerate. However, if we introduce LFV interactions, as it might be welcome in
order to account for the neutrino masses and oscillations, a significant mass splitting
$\Delta m_{\tilde{\ell}}/m_{\tilde{\ell}}$ can be still induced. In such a case,
$\Delta m_{\tilde{\ell}}/m_{\tilde{\ell}}$ will turn out to be related to low energy
LFV processes like $\ell_i\to\ell_j\gamma$.

We remind that, flavour mixings between the second and first slepton families are tightly
constrained by the non observation of ${\rm BR}(\mu\to e\gamma)$, hence, they cannot
typically induce testable values for $\Delta m_{\tilde{\ell}}/m_{\tilde{\ell}}$.
Still, as widely stressed in the literature~\cite{Hisano:2002iy}, such tight constraints
might be evaded thanks to cancellations among different contributions to ${\rm BR}(\mu\to e\gamma)$
when the main source of LFV arises from the RH slepton sector. An example of such situation
is studied in Ref.~\cite{Hisano:2002iy}, in the context of a non-universal Higgs mass
(NUHM) model. In that case, $\Delta m_{\tilde{\ell}}/m_{\tilde{\ell}}\sim 1\%$ is still
compatible with the present $\mu\to e\gamma$ bound, provided the SUSY input parameters
are tuned to get exact cancellations in ${\rm BR}(\mu\to e\gamma)$.

On the other hand, we would like to point out that sizable values for
$\Delta m_{\sle}/m_{\sle}$ can be more naturally generated through
LFV in the $\tau-\mu$ (or $\tau-e$) sector, as it might naturally arise from
the large mixing angle observed in atmospheric neutrino oscillation experiments.

In order to see this, let us consider for simplicity the illustrative case
of a CMSSM with a single source of LFV that we parameterize, as usual, by means
of the mass insertions (MIs) $(\delta_{\rm{XY}})_{ij}\equiv ({\tilde m}^{2}_{\rm{XY}})_{ij}/\sqrt{({\tilde m}^{2}_{\rm{XY}})_{ii}({\tilde m}^{2}_{\rm{XY}})_{jj}}$
where $i,j=1,3$ are flavour indices, $\rm X,Y = L,R$ refers to the chirality of the
corresponding SM fermions and $({\tilde m}^{2}_{\rm{XY}})_{ij}$ are the $3 \times 3$
blocks of the slepton mass matrix with given chirality $\rm{XY}$.

In such a case, we obtain two mass eigenstates that are a mixture of staus and smuons such that $m^{2}_{\sle_{1,2}}= m^{2}_{\sle}(1 \mp \delta_{32})$, where for the moment we are neglecting
the effect of the LR stau mixing term.

Given that in the absence of LFV sources all the three slepton generations were degenerate, a mass splitting between the third and the second generations will clearly imply also a mass splitting between the first two slepton generations, $\Delta m_{\tilde{\ell}}/m_{\tilde{\ell}}$.

It is straightforward to check that, in terms of the mass insertion $\delta_{32}$, such mass
splitting is approximately given by
\beq
\left|\frac{\Delta m_{\tilde{\ell}}}{m_{\tilde{\ell}}}\right|
\simeq \frac{|\delta_{32}|}{2}\,.
\label{dm_vs_d32}
\eeq
Hereafter, we denote both the mass splitting between $\tilde{e}_L-\tilde{\mu}_L$ and
$\tilde{e}_R-\tilde{\mu}_R$ (induced by $(\delta_{\rm LL})_{32}$ and $(\delta_{\rm RR})_{32}$, respectively) with $\Delta m_{\tilde{\ell}}$.

The mass splitting of Eq.~(\ref{dm_vs_d32}) might be measured by the LHC with an accuracy level
better than the percent~\cite{Allanach:2008ib}, hence, such a measurement would enable us to
test the presence of LFV effects in some classes of SUSY theories, by detecting flavour conserving
processes such as $\nt_{2}\to\nt_{1}e^{+}e^{-}$, $\nt_{2}\to\nt_{1}\mu^{+}\mu^{-}$.

In passing, we notice that LFV sources in the $\tau-\mu$ sector clearly induce also a mass
splitting between the third and second slepton generations. Therefore, in principle, one
could make use of the edge splitting measurements for the $\tau-\tau$ and $\mu-\mu$ invariant
mass distributions to extract information about the LFV source $\delta_{32}$.
However, in practice, this could be hard because i) experimentally the edge measurement for
the $\tau-\tau$ case is more challenging than the electron and muon ones, ii) the staus and
the smuons are split even in the absence of LFV due to both RGE effects driven by the
Yukawa interactions ($\propto y_\tau^2$) and the presence of a left-right mixing term that
affects mostly the stau masses. As a result, the measurement of the edge splitting for the
$\mu-\mu$ and $e-e$ invariant mass distributions seems to be the most suitable tool to unveil
LFV effects.

Yet, it should be stressed that a mass splitting between smuons and selectrons does not
necessarily represent a clear probe of LFV effects. In fact, one could always envisage
a situation where the lepton and slepton mass matrices are enough aligned \cite{Nir:1993mx}, 
in order to avoid LFV effects, while the masses for different slepton generations are non 
degenerate. The measurement of the slepton mass splittings in this latter case  
have been studied in Ref.~\cite{Nir}.

Hence, in order to test whether the slepton mass splittings come from LFV sources,
it is crucial to have direct signals from LFV processes as we are going to 
discuss in the following section.

\section{LFV at the LHC and at low-energy experiments}\label{Sec:3}

As discussed in the previous section, LFV sources for the $\tau-\mu$ transition
naturally induce a smuon/selectron splitting that could be detected by means of
edge splittings for the $\mu-\mu$ and $e-e$ invariant mass distributions of the
processes $\nt_{2}\to\nt_{1}\mu^{\pm}\mu^{\mp}$ and $\nt_{2}\to\nt_{1}e^{\pm}e^{\mp}$.
Obviously, at the same time, also the LFV process $\nt_{2}\to\nt_{1}\tau^{\pm}\mu^{\mp}$,
that represents one of the most promising LFV decay mode at the LHC~\cite{hinchliffe,ellis,agashe,other},
is generated.

In fact, the decay of $\nt_2$ in a lepton and slepton, followed by the decay of the slepton
in a different lepton and the LSP (${\rm BR}(\nt_2 \to \ell_i \ell_j \nt_1)$), is a tree-level process, 
which simply depends on the misalignment between the slepton and lepton mass eigenstates \cite{arkani}. 
Furthermore, there is no `GIM-like' suppression ($\sim \Delta m_{\sle}/\bar{m}_{\sle}$ in the amplitude)
in the case of the intermediate sleptons being real ($m_{\nt_2}>m_{\sle_\alpha}$).

Unfortunately, the broad parameter space in the CMSSM which can be probed by the LHC is
already excluded by the $\mu\to e\gamma$ constraint when there is a LFV sources in the
$\mu-e$ sector~\cite{Hisano:2002iy} (barring fine tuned cases where accidental
cancellations strongly reduce ${\rm BR}(\mu\to e\gamma)$~\cite{Hisano:2002iy}).

Therefore, in the following, as mentioned above, we consider a possible mixing between
$\tau$- and $\mu$ sleptons (both left-handed and right-handed), since the constraints
for LFV in the $\tau-\mu$ sector are less stringent than in the $\mu-e$ sector.

In principle, we could also consider the case of ${\tilde\tau}-{\tilde e}$ mixing. However, the possibility of having {\it simultaneously} large effects in $\nt_{2}\to\nt_{1}\tau^{\pm}\mu^{\mp}$
and $\nt_{2}\to\nt_{1}\tau^{\pm} e^{\mp}$, implying an effective $\mu-e$ mixing 
$\delta_{\mu e}\sim \delta_{\mu\tau}\delta_{\tau e}$, is again strongly constrained by the bounds
on ${\rm BR}(\mu\to e\gamma)$~\cite{Paradisi:2005fk,Ibarra:2008uv}.

Therefore, neglecting the contribution of virtual intermediate sleptons,
the branching ratio of the LFV neutralino decay can be written as:
\bea\label{eq:8}
{\rm BR}(\nt_2 \to \ell_i \ell_j \nt_1) =\!\!\!\!\!\!
&&\bigg[
{\rm BR}(\nt_2 \to \ell_i \sle_\alpha){\rm BR}(\sle_\alpha \to \ell_j \nt_1) +
\nonumber\\
&&{\rm BR}(\nt_2 \to \ell_j \sle_\alpha){\rm BR}(\sle_\alpha \to \ell_i \nt_1)
\bigg]
\eea
where the sum is understood over the sleptons lighter than the neutralino.
The relevant decay widths are given by~\cite{widths}:
\begin{align}
\Gamma(\nt_K\!\to\!\sle_\alpha \ell_i )\! = &{\alpha_2 \over 16} m_{\nt_K}
\!\left(\!1\!-\!{m^2_{\sle_\alpha} \over m^2_{\nt_K} } \!\right)^2\!
\!\left(\left|L^K_{i\alpha}\right|^2 \!+\! \left|R^K_{i\alpha}\right|^2 \right)
\nonumber\\
\Gamma( \sle_\alpha\!\to\!\nt_K \ell_i ) \!= &{\alpha_2 \over 8} m_{\sle_\alpha}\!
\!\left(\!1\!-\!{m^2_{\nt_K}\over m^2_{\sle_\alpha}} \!\right)^2\!
\!\left(\left|L^K_{i\alpha}\right|^2 \!+\! \left|R^K_{i\alpha}\right|^2 \right)
\label{eq:widths}
\end{align}
where the masses of ordinary leptons have been neglected, $K=1,4$, $\alpha=1,6$ and
$i=1,3$ ($l_i = (e,\mu,\tau)$) and $L^i_{K\alpha}$, $ R^i_{K\alpha}$ represent the lepton-slepton-neutralino interaction vertices:
\begin{align}
 & L^K_{i\alpha} \!=\! -\left[ N_{K2} \!+\! N_{K1} \tan\theta_W \right] U_{\alpha i}
  + {m_{l_i} \over M_W \cos\beta} N_{K 3} U_{\alpha (i+3)} \nonumber\\
&  R^K_{i\alpha} =  2 N_{K1} U_{\alpha (i+3)} \tan\theta_W
  + {m_{l_i} \over M_W \cos\beta} N_{K 3} U_{\alpha i} \,.
\label{eq:vertices}
\end{align}
Here $U$ and $N$ are the matrices which rotate sleptons and neutralinos
into their mass eigenstates.

In order to compute the total $\nt_2$ width, and then 
${\rm BR}(\nt_{2} \to \tilde{\ell}_{\alpha} \ell_{j} \to \nt_{1}\ell_{i}^{\pm}\ell_{j}^{\mp})$, 
one has to consider the following flavour-violating and -conserving $\nt_{2}$ decays:
\globallabel{3h}
\begin{align}
\nt_{2} & \to \tilde{\ell}_{\alpha} \ell_{j} \to \nt_{1}\ell_{i}^{\pm}\ell_{j}^{\mp}
\mytag \\
\nt_{2} & \to \nt_1 Z^0 \to \nt_1 \ell^{+}_{i} \ell^{-}_{i}
\mytag \\
\nt_{2} & \to \nt_1 h^0 \to \nt_1 \ell^{+}_{i} \ell^{-}_{i} \mytag \\
\nt_{2} & \to \tilde{\nu}_{\alpha} \nu_{j} \to \nt_{1} \nu_{i} \nu_{j}
\mytag 
\end{align}
Clearly, the $h^0$ decays may provide a sizable rate only to the flavour-conserving decay
$\nt_{2}\to\nt_{1}\tau^{\pm}_{i}\tau^{\mp}_{i}$.
Decays into squark-quark are not taken into account, since $\nt_2$ is always lighter than
squarks in the SUSY framework considered here. 

The presence of $\tau-\mu$ LFV sources will also induce, at one loop, LFV low energy processes
like $\tau\to\mu\gamma$ and so on. In table~\ref{tab:lfvtable}, we report the current and expected experimental bounds on some of the $\tau-\mu$ transitions.

\begin{table}[t]
\addtolength{\arraycolsep}{3pt}
\renewcommand{\arraystretch}{1.3}
\centering
\begin{tabular}{|l|c|c|r|}
\hline
Process & Present Bound & Future Bound  & Future Exp.\\
\hline\hline
BR($\tau \to \mu\,\gamma$) & $4.4 \times 10^{-8}$ & $\mathcal{O}(10^{-8}) $ &SuperB \cite{KEK}
\\
BR($\tau \to \mu\, \mu\, \mu$) & $3.2 \times 10^{-8}$ & $\mathcal{O}(10^{-8}) $ &LHCb \cite{LHCb}
\\
BR($\tau \to \mu\, e\,e$) & $2.0 \times 10^{-8}$ & $\mathcal{O}(10^{-8}) $ &SuperB \cite{KEK}
\\
\hline
\end{tabular}
\caption{\small
Present \cite{present-bounds} and upcoming experimental limits on various $\tau-\mu$
transitions at 90\% C.L.}
\label{tab:lfvtable}
\end{table}

The branching ratio of $\tau\to\mu\gamma$ can be written as
\bea
\frac{{\rm BR}(\tau\to\mu\gamma)}{{\rm BR}(\tau\to\mu\nu_{\tau}\bar{\nu_{\mu}})} =\frac{48\pi^{3}\alpha}{G_{F}^{2}}(|A_L^{32}|^2+|A_R^{32}|^2)\,.
\eea

Although in the numerical analysis we perform a full computation of LFV processes using the exact formulae of Ref.~\cite{Hisano:1995cp}, we report, in the following, the amplitudes as obtained in
the MI approximation, within the illustrative case of a degenerate SUSY spectrum with a common
mass ${\tilde m}$~\cite{Paradisi:2005fk}:
\begin{align}
  A^{32}_L
  &\simeq
  \frac{\alpha_2}{60\pi}\frac{\tan\beta}{\tilde{m}^{2}}
  (\delta_{\rm LL})_{32}\,, \nonumber
  \\
  A^{32}_R
  &\simeq
  -\frac{\alpha_1}{4\pi}\frac{\tan\beta}{\tilde{m}^{2}}
  \frac{(\delta_{\rm RR})_{32}}{60}\,. 
\label{MI_degenerate}
\end{align}
As we will discuss in detail in the next sections, $\tau\to\mu\gamma$ and
$\nt_{2}\to\nt_{1}\tau^{\pm}\mu^{\mp}$ provide complementary probes of the
SUSY parameter space, especially when combined also with the possible measurement
of the edge splittings. First, we notice that $\rm{BR}(\tau\to\mu\gamma)$ scales 
as $\tan^{2}\beta$ while, $\rm{BR}(\nt_{2}\to\nt_{1}\tau^{\pm}\mu^{\mp})$ is
sensitive to $\tan\beta$ for kinematical reasons, even though it does not depend
explicitly on it.
Second, ${\rm BR}(\tau\to\mu\gamma)\sim \tilde{m}^{-4}$ and thus it decouples fast
with the SUSY scale, in contrast to $\rm{BR}(\nt_{2}\to\nt_{1}\tau^{\pm}\mu^{\mp})$.

\begin{figure}[t]
\includegraphics[scale=0.5, angle=-90]{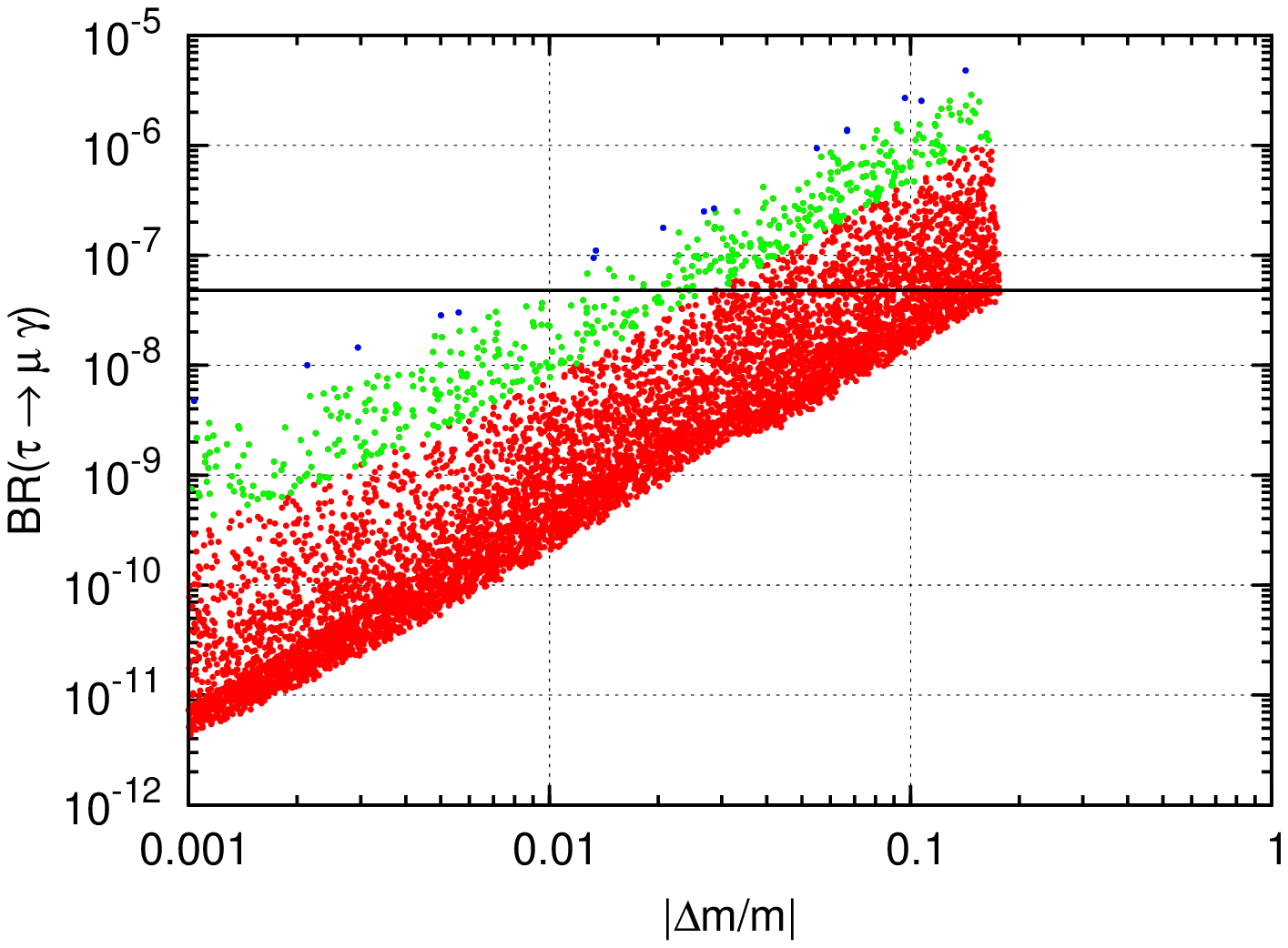}
\includegraphics[scale=0.5, angle=-90]{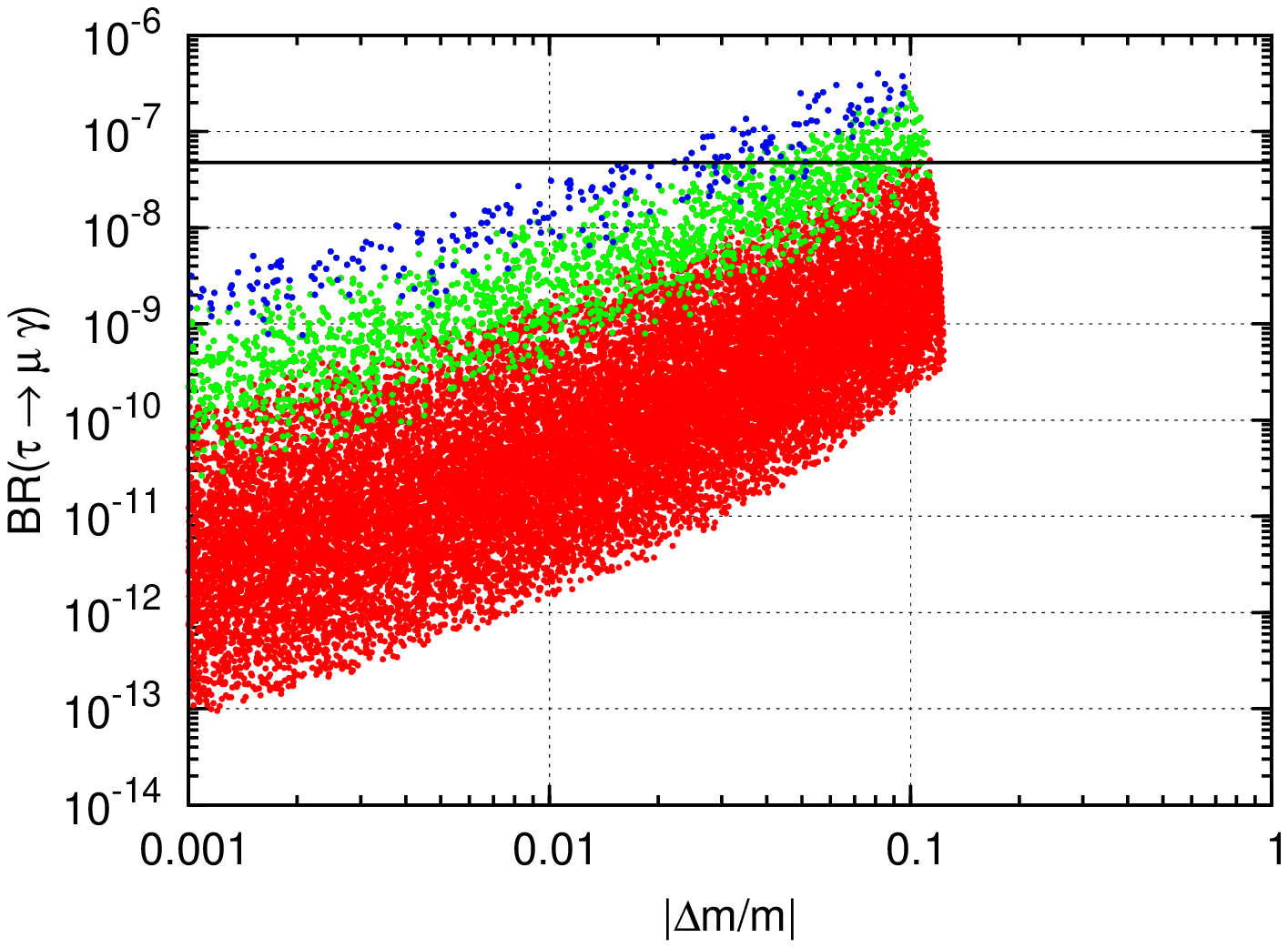}
\caption{Top: BR$(\tau\to\mu\gamma)$ vs. $\Delta m_{\sle}/m_{\sle}$
in case of LFV sources in the left-left slepton sector.
Bottom: the same for the right-right slepton sector.}
\label{scatter}
\end{figure}

\section{Numerical analysis}\label{Sec:4}

In this section, we present our numerical results, obtained in the framework
of the CMSSM in presence of LFV sources. In order to compute the SUSY spectrum,
we numerically solved the full 1-loop RGEs of the MSSM, switching on LFV mass-insertions
at low-energy. Then, we compute the relevant processes by means of a full calculation
in the mass eigenstate basis. For each point of the parameter space, we impose 
the following constraints: (i) successful EWSB and absence of tachyonic particles;
(ii) limits on SUSY masses from direct searches, (iii) all the 
currently available hadronic flavour constraints \cite{ABGPS}.

In Fig.~\ref{scatter}, we plot BR$(\tau\to\mu\gamma)$ as a function of
$\Delta m_{\sle}/m_{\sle}$, for the following choice of the SUSY
parameters: $\tan\beta=10$, $0 < m_0 < 1000$ GeV, $0 < M_{1/2} < 1000$ GeV, $A_0=0$.
In the upper plot, we switch on only the LFV MI $(\delta_{\rm LL})_{32}$ and
we vary it in the range $10^{-3}<(\delta_{\rm LL})_{32}<0.3$. All the points
of Fig.~\ref{scatter} are such that the processes $\nt_{2}\to\nt_{1}e^{+}e^{-}$
and $\nt_{2}\to\nt_{1}\mu^{+}\mu^{-}$ through real sleptons are kinematically
allowed, namely they satisfy $m_{\tilde{e}_{L}},\,m_{\tilde{\mu}_{L}} < m_{\nt_2}$.
In the lower plot of Fig.~\ref{scatter}, we scan $(\delta_{\rm RR})_{32}$
also in the range $10^{-3}<(\delta_{\rm RR})_{32}<0.3$ and we require that 
$m_{\tilde{e}_{R}},\,m_{\tilde{\mu}_{R}} < m_{\nt_2}$.

The grey points give $a^{\rm SUSY}_\mu\equiv(g-2)^{\rm SUSY}_\mu/2 > 10^{-9}$,
while for the blue points $a^{\rm SUSY}_\mu> 2\times 10^{-9}$. The black horizontal
line represents the present bound BR$(\tau\to\mu\gamma)< 4.4\times 10^{-8}$.

We observe that, in the case of left-handed sleptons, mass splittings of order $\lesssim 3\%$,
are compatible with a solution of the $(g-2)_{\mu}$ anomaly at the 2-$\sigma$ level, i.e.
$a^{\rm SUSY}_\mu> 10^{-9}$, while satisfying at the same time the $\tau\to\mu\gamma$ bound.
Moreover, in the case of right-handed sleptons, the mass splitting can reach even the
10 $\%$ level.

We have also found that, in the same ranges for the SUSY parameters as in Fig.~\ref{scatter},
the edge splitting ${\Delta m_{ll}}/{m_{ll}}$ is unambiguously enhanced (at least by a factor
of $\sim 3$) with respect to the slepton mass splitting $\Delta m_{\sle}/m_{\sle}$
when the LFV arises in the LL sector. In contrast, for LFV sources in the RR sector,
${\Delta m_{ll}}/{m_{ll}}$ can be both enhanced or suppressed (up to a factor of $\sim 3$) 
compared to $\Delta m_{\sle}/m_{\sle}$.
\begin{figure}[t]
\includegraphics[scale=0.45, angle=-90]{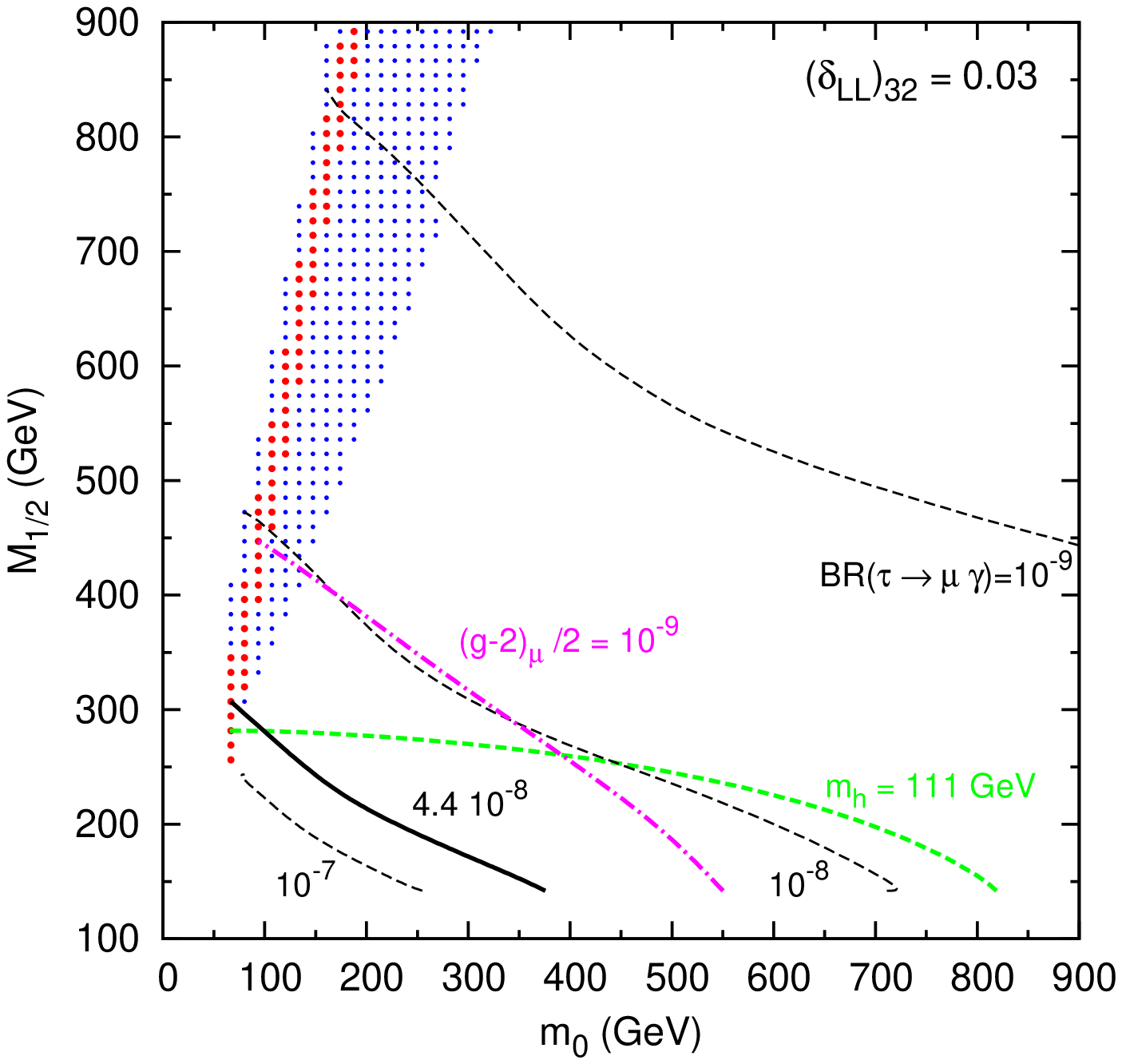}
\includegraphics[scale=0.45, angle=-90]{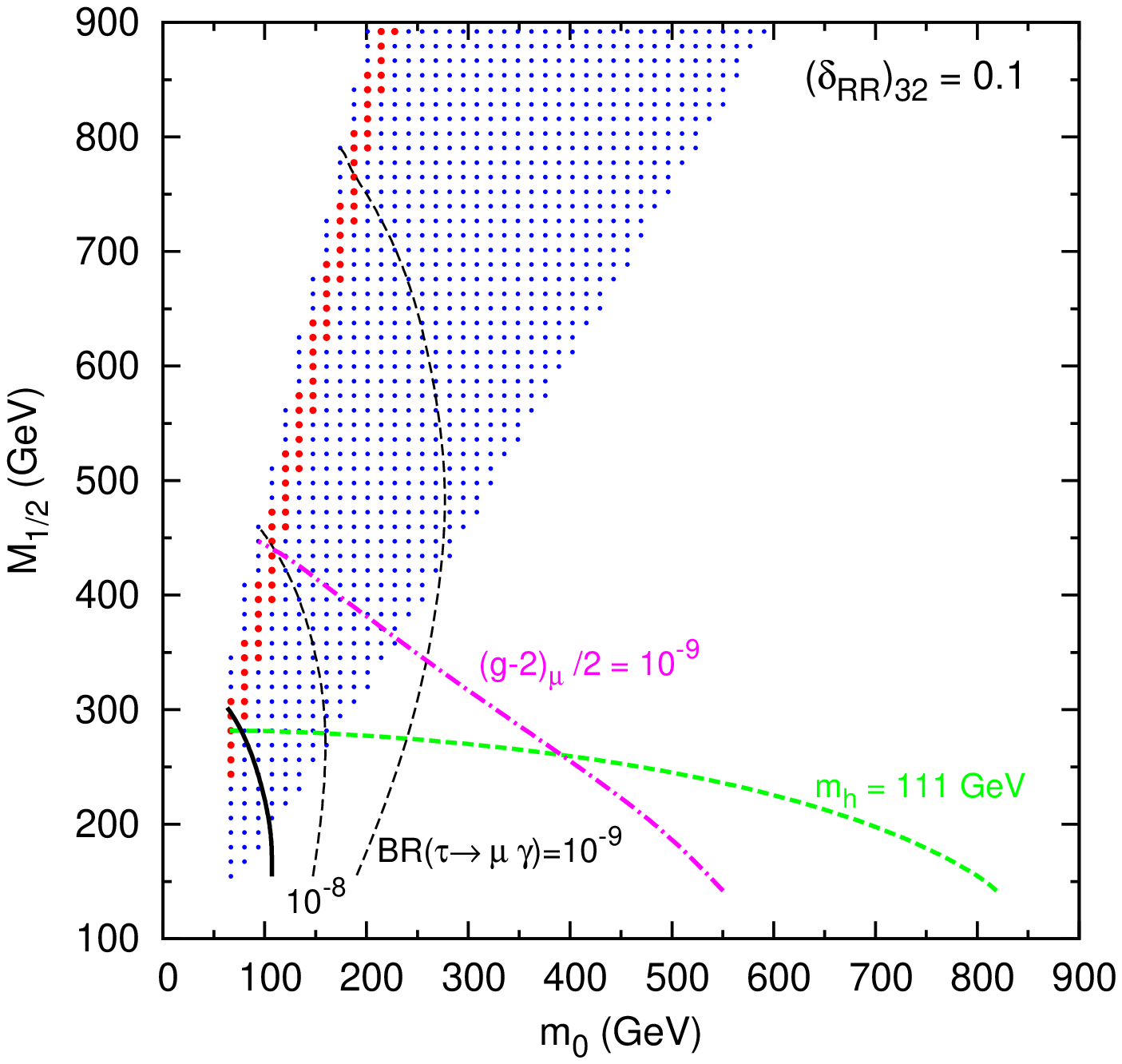}
\caption{Predictions for $\rm{BR}(\tau\to\mu\gamma)$ in the $(m_0,\,M_{1/2})$
plane for $\tan\beta =10$, $A_0=0$, $(\delta_{\rm LL})_{32} = 0.03$ (top) and
$(\delta_{\rm RR})_{32} = 0.1$ (bottom). In both plots, the process $\tilde{\chi}^{0}_{2}\to\tilde{\chi}^{0}_{1}\ell\ell$ (mediated by a real slepton)
is kinematically allowed in the region marked with blue dots. The red points
additionally satisfy the Dark Matter constraints.}
\label{par-space}
\end{figure}
\begin{figure}[t]
\includegraphics[scale=0.45, angle=-90]{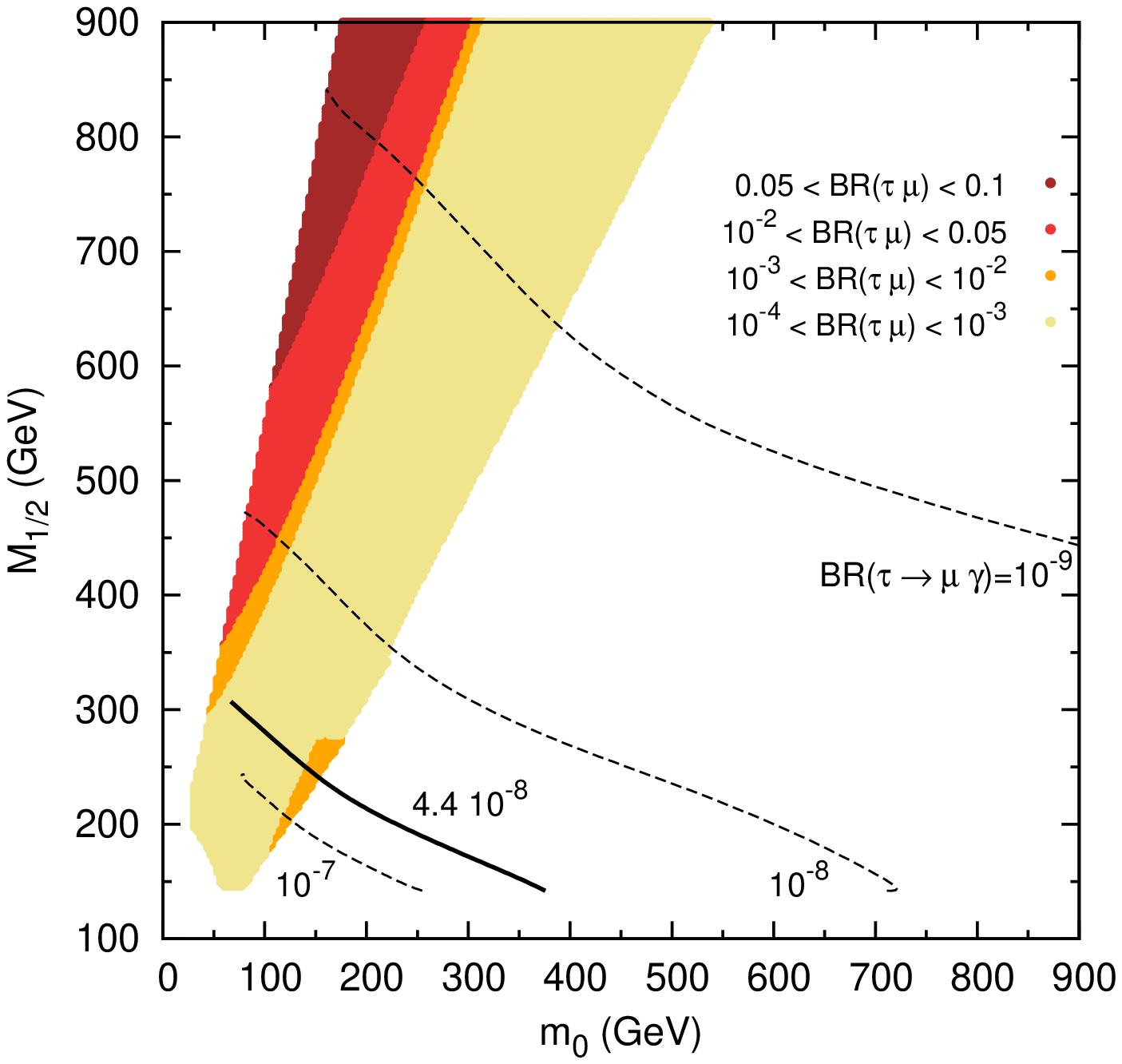}
\includegraphics[scale=0.45, angle=-90]{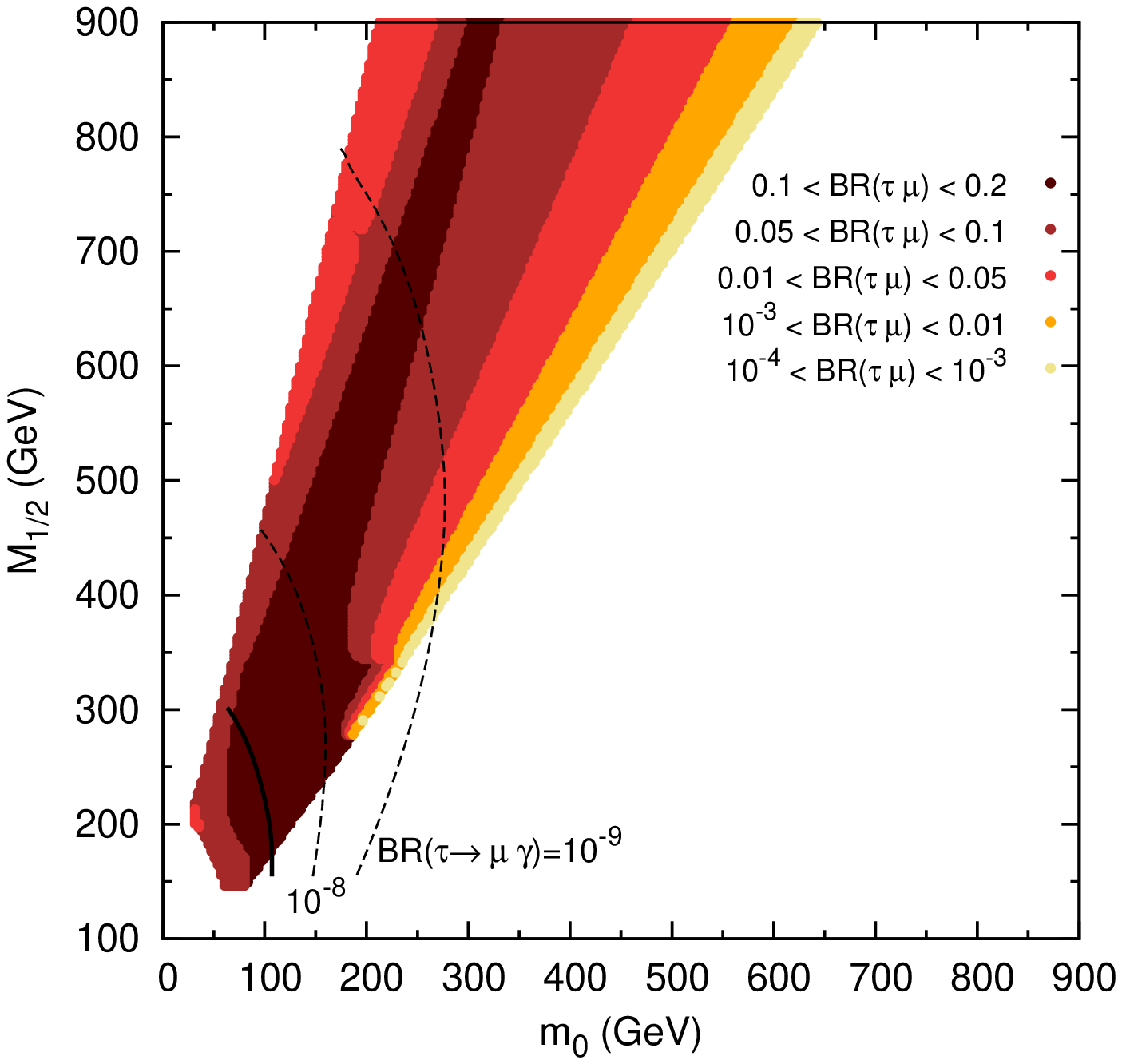}
\caption{Predictions for $\rm{BR}(\nt_{2}\to\nt_{1}\tau^{\pm}\mu^{\mp})$ and
$\rm{BR}(\tau\to\mu\gamma)$ in the $(m_0,\,M_{1/2})$ plane for $\tan\beta =10$,
$A_0=0$, $(\delta_{\rm LL})_{32} = 0.03$ (top) and $(\delta_{\rm RR})_{32} = 0.1$
(bottom).
The $\rm{BR}(\nt_{2}\to\nt_{1}\tau^{\pm}\mu^{\mp})$ decreases passing from the
darker to the lighter regions.}
\label{par-space-chi2chi1tm}
\end{figure}

Since a precise measurement of the slepton masses would require $m_\sle < m_{\nt_2}$, it is
useful to display the values of $m_0$ and $M_{1/2}$ accounting for $m_\sle < m_{\nt_2}$.

In Fig.~\ref{par-space}, we show the $(m_0,\,M_{1/2})$ plane for $\tan\beta =10$, $A_0=0$,
$(\delta_{\rm LL})_{32} = 0.03$ (top) and $(\delta_{\rm RR})_{32} = 0.1$ (bottom).
We chose the above values of the LFV sources as illustrative cases of scenarios
where ${\rm BR}(\tau\to\mu\gamma)$ is kept under control and, at the same time,
sizable SUSY contributions to $(g-2)_{\mu}$ are still possible, as shown by Fig.~\ref{scatter}.

The region of Fig.~\ref{par-space} marked with blue dots corresponds to the kinematically
allowed region for the decays of neutralinos into smuons and selectrons.
Their branching fractions here are always above the percent level ($\gtrsim 3\%$) in
the case of $(\delta_{\rm LL})_{32}=0.03$, while, for $(\delta_{\rm RR})_{32}=0.1$, they
are around 1\% in the region favoured by $(g-2)_{\mu}$ and get lower outside it.
This region includes also the neutralino-stau coannihilation strip (red dots).

In the upper plot, the MI $(\delta_{\rm LL})_{32}=0.03$ induces a mass splitting
$(\Delta m_{\sle}/m_{\sle})_L$ around 1-1.5 $\%$. In the lower plot, the MI
$(\delta_{\rm RR})_{32}=0.1$ gives $2\%\lesssim(\Delta m_{\sle}/m_{\sle})_R\lesssim 4\%$.
In both figures, the black thick line refers to the current bound on BR$(\tau\to\mu\gamma)$,
the purple dashed-dot line corresponds to $a^{\rm SUSY}_\mu = 1\times10^{-9}$ while the green
dashed line accounts for the lightest Higgs boson bound (we impose $m_h>111$ GeV taking into
account a theoretical uncertainty of 3 GeV).
\begin{figure}[t]
\includegraphics[scale=0.5, angle=-90]{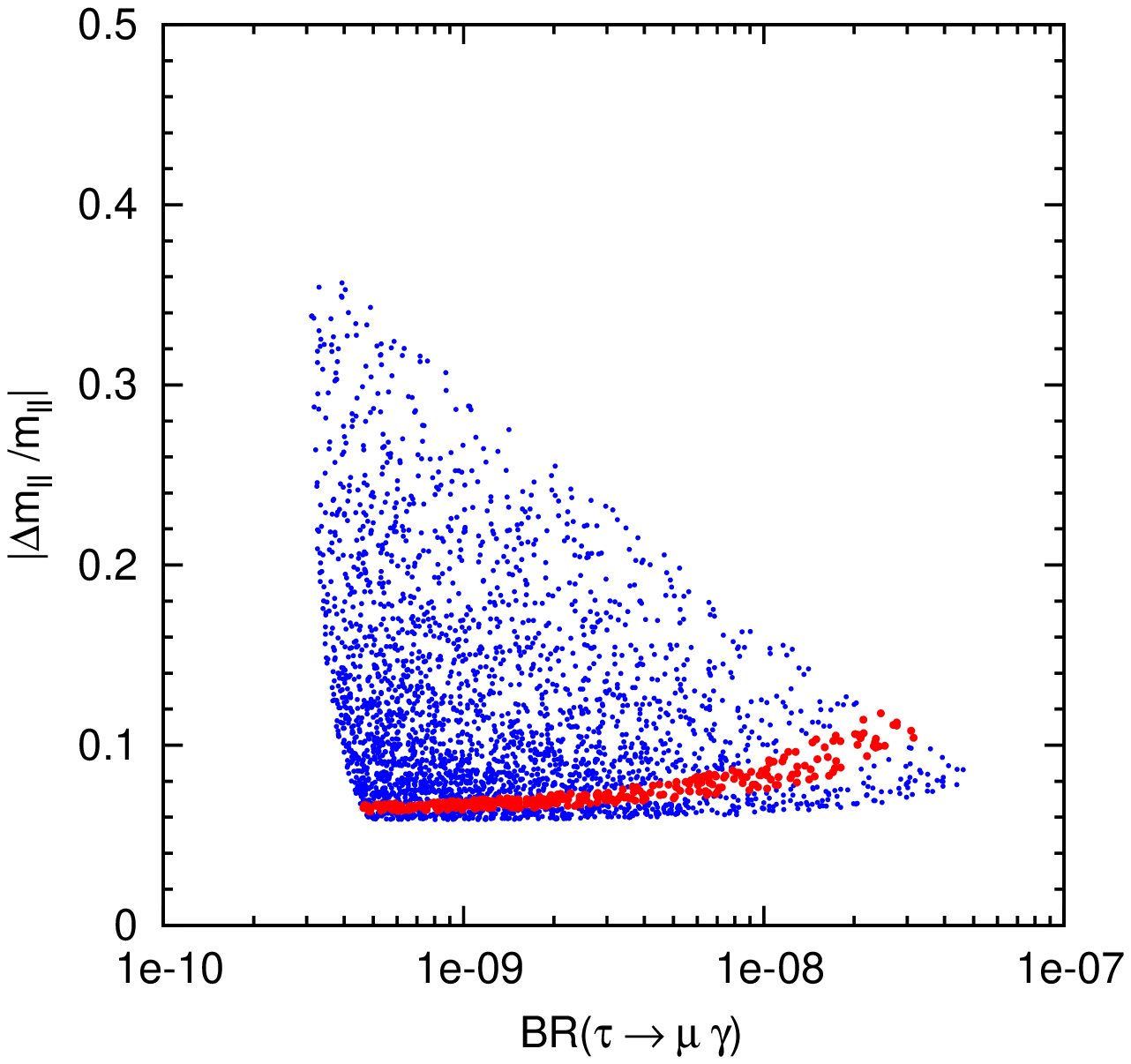}
\includegraphics[scale=0.5, angle=-90]{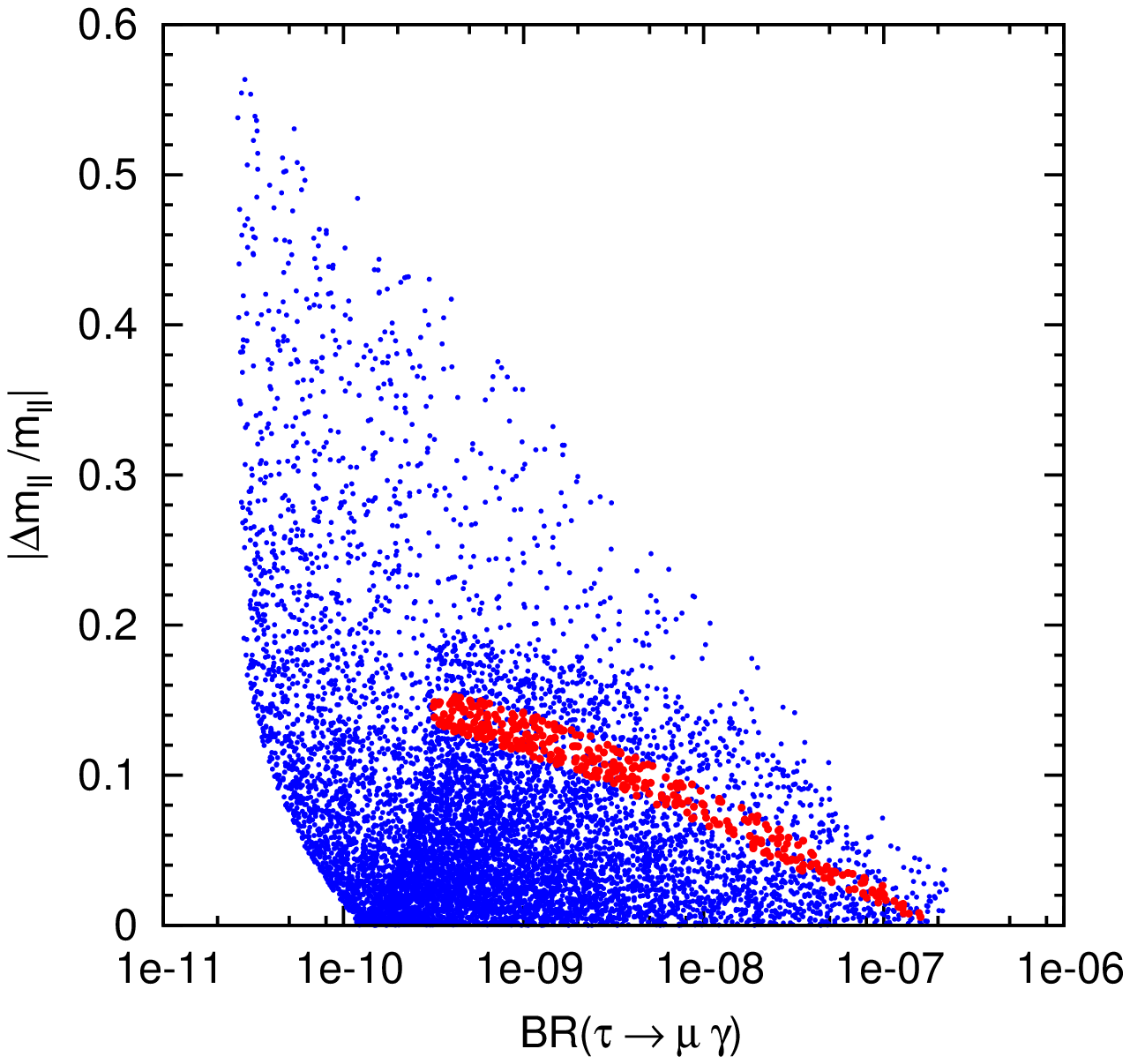}
\caption{$|{\Delta m_{ll}}/{m_{ll}}|$ vs. BR$(\tau\to\mu\gamma)$ for the points
corresponding to the blue and red regions of Fig.~\ref{par-space}.}
\label{edgeplot2}
\end{figure}

Interestingly enough, both plots of Fig.~\ref{par-space} show that there are sizable
regions of the parameter space where, at the same time: (i) BR$(\tau\to\mu\gamma)$
satisfies the current experimental bound while being within the reach of a SuperB factory
at KEK (BR$(\tau\to\mu\gamma)>10^{-8}$ \cite{KEK}), (ii) the $(g-2)_{\mu}$ anomaly can be 
explained at the 2-$\sigma$ level, i.e. $a^{\rm SUSY}_{\mu}\gtrsim 1\times 10^{-9}$;
(iii) the WMAP relic density constraint can be fulfilled by an effective neutralino-stau
coannihilation; (iv) selectrons and smuons can be produced by cascade decays through the
next-to-lightest neutralino.

In Fig.~\ref{par-space-chi2chi1tm}, we show the predictions for 
$\rm{BR}(\nt_{2}\to\nt_{1}\tau^{\pm}\mu^{\mp})$ and $\rm{BR}(\tau\to\mu\gamma)$ in the 
$(m_0,\,M_{1/2})$ plane for the same input parameters as in Fig.~\ref{par-space}. 
We notice that the kinematically allowed region of Fig.~\ref{par-space-chi2chi1tm} is 
significantly larger than that of Fig.~\ref{par-space}. The reason can be traced back 
remembering that the staus are lighter than $\nt_{2}$ in a broader region of the SUSY 
parameter space compared to the selectrons and smuons.
 In these plots, $\rm{BR}(\nt_{2}\to\nt_{1}\tau^{\pm}\mu^{\mp})$ decreases passing
from the darker to the lighter regions.

Interestingly enough, Fig.~\ref{par-space-chi2chi1tm} clearly shows that 
$\rm{BR}(\nt_{2}\to\nt_{1}\tau^{\pm}\mu^{\mp})$ and $\rm{BR}(\tau\to\mu\gamma)$
are complementary probes of LFV in SUSY with $\rm{BR}(\nt_{2}\to\nt_{1}\tau^{\pm}\mu^{\mp})$
being potentially more powerful than $\rm{BR}(\tau\to\mu\gamma)$ especially for a heavy 
SUSY spectrum.

As mentioned above, a smuon/selectron mass splitting at the percent level would be potentially
measurable at the LHC especially if the edge splitting ${\Delta m_{ll}}/{m_{ll}}$ receives an 
enhancement. In order to check this crucial point, in Fig.~\ref{edgeplot2} we plot 
$|{\Delta m_{ll}}/{m_{ll}}|$ (as generated by LFV effects) vs. BR$(\tau\to\mu\gamma)$, selecting 
the points corresponding to the blue-dotted region of Fig.~\ref{par-space}. We impose on those 
points the additional requirements $(m_{\nt_2} - m_{{\tilde e},~{\tilde \mu}}) \geq 10$ GeV,
$(m_{{\tilde e},~{\tilde \mu}}-m_{\nt_1}) \geq 10$ GeV, so that the leptons are not too soft for 
detection. 

As shown by Fig.~\ref{edgeplot2}, ${\Delta m_{ll}}/{m_{ll}}$ might be quite large, well beyond 
the 10 $\%$ level, even for BR$(\tau\to\mu\gamma)\lesssim 10^{-9}$ thus beyond the reach of a 
Super Flavour Factory \cite{super-flavour}.

Focusing on the coannihilation strip, corresponding to the red points of Fig.~\ref{edgeplot2},
we observe that, in the case of $(\delta_{\rm LL})_{32}=0.03$, the splitting of the $\mu-\mu$ and
$e-e$ edges (coming from cascade decays through $\tilde{\mu}_L$ and $\tilde{e}_L$, respectively)
is around 8 $\%$.
For $(\delta_{\rm RR})_{32} = 0.1$, the edge splitting (now corresponding to intermediate
$\tilde{\mu}_R$ and $\tilde{e}_R$) can be of order 10$\%$ or more.

The above edge splittings, that are well measurable at the LHC, can probe the SUSY parameter
space in regions where BR$(\tau\to\mu\gamma)$ is experimentally challenging or even unreachable.

\section{Production cross-sections, signals and backgrounds}\label{Sec:5}

In the following, we are going to estimate  the cross-sections, signals and backgrounds
for the relevant processes discussed in the previous sections, in order to provide information
about the capability of the LHC of detecting such processes and, in particular, of measuring
the slepton masses thanks to the kinematical end-points technique.

A full simulation of the production, decays and detection of SUSY particles at the LHC is
beyond the scope of the present paper. Nevertheless, we can get a semi-quantitative idea of
the size of the cross-sections, signals and backgrounds, by using some approximate
formulae provided in the literature.

\subsection{Production cross-sections}
At the LHC, operating at 14 TeV, the total production cross-section for squarks and gluinos,
within the CMSSM, can be estimated to be \cite{hinchliffe}:
\begin{equation}
 \sigma_{\rm SUSY} \simeq 1.79\times 10^{13} (0.1 m_0 + M_{1/2})^{-4.8} ~{\rm pb}\,.
\end{equation}
This formula gives an agreement with the full computation to within 25\% \cite{hinchliffe,Hisano:2002iy}.

In order to estimate the number of (Wino-like) $\nt_2$ produced from cascade of squarks and gluinos,
we will take ${\rm BR}({\tilde q}_L \to q_L\, \nt_2)\simeq 1/3$ and assume the probability to produce
a ${\tilde q}_L$ to be 50\%. Hence, the production cross-section of $\nt_2$ can be estimated to be:
\begin{equation}
 \sigma_{\nt_2} \simeq \frac{1}{2}~ \sigma_{\rm SUSY} \times{\rm BR}({\tilde q}_L \to q_L\, \nt_2) 
 \simeq \frac{1}{6}~\sigma_{\rm SUSY}\,.
\end{equation}

Making extensive use of the formulae given above and computing numerically the branching fractions 
of the second neutralino decays, we are going to estimate the following quantities:
\globallabel{eq:cross-sections}
\begin{align}
 \sigma_{ee}\! &\equiv \! \sigma(\nt_2 \!\to\! \nt_1 e^+ e^-) \!= \!\sigma_{\nt_2} \!\times\! {\rm BR}(\nt_2 \!\to\! \nt_1 e^+e^-)\mytag \\
 \sigma_{\mu \mu}\! &\equiv \! \sigma(\nt_2 \!\to\! \nt_1 \mu^+ \mu^-) \!= \!\sigma_{\nt_2} \!\!\times\! {\rm BR}(\nt_2 \!\to\! \nt_1 \mu^+ \mu^- \!) \mytag\\
 \sigma_{\tau \mu}\! &\equiv \! \sigma(\nt_2 \!\to\! \nt_1 \tau^\pm \mu^\mp) \!= \!\sigma_{\nt_2} \!\times\! {\rm BR}(\nt_2 \!\to\! \nt_1 \tau^\pm \mu^\mp) \mytag
\end{align}

\subsection{Signals and backgrounds}
Let us now estimate signals and relevant backgrounds following closely the analysis
of Ref.~\cite{agashe}). We are considering opposite sign dilepton events (flavour
conserving and flavour violating ones), which can have both SM and SUSY backgrounds.

The SM background mainly comes from $W^+W^-$ and $t\bar{t}$ production. Both these
sources can be strongly reduced by imposing the standard cuts (on missing $p_T$,
number of jets, jets $p_T$, etc.) used to discriminate SUSY events from the SM
background (see for instance the discussion in Ref.~\cite{hinchliffe,ellis}).
We will therefore concentrate only on the SUSY background assuming that the SM one
can be sufficiently reduced.

We will parameterize the effect of the above mentioned kinematic cuts by means
of an acceptance factor $\epsilon_{\rm cut}$, affecting both our signals and the
corresponding SUSY background.

Let us start considering the opposite sign same flavour events we want to study
for extracting the slepton masses. The signal is given by:
\begin{equation}
S_{\ell^+\ell^-} = \sigma_{\ell\ell}
\times\epsilon_{\ell}^2\times \epsilon_{\rm cut} 
\times L\,,
\label{eq:sll}
\end{equation}
where $\epsilon_{\ell}$ is the lepton detection efficiency and $L$ the integrated luminosity.

There are two main SUSY sources of background to these events. The first one comes from pairs
of left-handed squark-antisquark, when both the particles decay into opposite sign charginos:
\begin{equation}
{\tilde q}_L {\tilde q}_L^* \to \ch^+_1 \ch^-_1 + \cdots
\label{eq:charginos}
\end{equation}
The charginos can then produce opposite sign leptons (either with same or different flavours) 
by decaying as follows:
\begin{align*}
\ch^\pm_1 & \to \tilde{\nu} \,\ell^\pm \,, \\
\ch^\pm_1 & \to \sle^\pm \, \nu \, \to \, \ell^\pm \, \nu \, \nt \,, \\
\ch^\pm_1 & \to W^\pm \, \nt \, \to \, \ell^\pm \, \nu \, \nt \,.
\end{align*}
The number of such background events can be estimated to be:
\begin{eqnarray}
B_{\ell^+ \ell^-}^{\ch^+\ch^-}&=&
\sigma_{\ch^+\ch^-}\times\epsilon_{\ell}^2\times\epsilon_{\rm cut}\times L \times
\nonumber \\
\!\!&\times&\!\!
\bigg[
{\rm BR}(\ch^\pm_1 \!\to\! \tilde{\nu} \ell^\pm) + \nonumber \\
\!\!&+&\!\!
{\rm BR}(\ch^\pm_1 \!\to\! \sle^\pm \nu)\,{\rm BR}(\sle^\pm \!\to\! \ell^\pm \nt) + \nonumber \\
\!\!&+&\!\!
{\rm BR}(\ch^\pm_1 \!\to\! W^\pm \nt)\,{\rm BR}(W^\pm \!\to\! \ell^\pm \nu)
\bigg]^2 \!,
\label{eq:bg1}
\end{eqnarray}
where $\sigma_{\ch^+\ch^-}$ is the production cross-section for the pair $\ch^+_1\ch^-_1$
from the squarks decay of Eq.~(\ref{eq:charginos}). Using public packages for computing
the SUSY production cross-sections \cite{prospino} and decays \cite{susyhit}, we found
that $\sigma_{\ch^+\ch^-} \simeq \left(\frac{1}{3}\div\frac{1}{2}\right) \sigma_{\nt_2}$
for all the points A-D we are going to study in the next section.

Since the angle between the two leptons $\theta_{\ell^+\ell^-}$ from the decays of $\ch^+_1\ch^-_1$ 
is likely to be larger than in the case of the decay of a single $\nt_2$, the background 
of Eq.~(\ref{eq:bg1}) can be reduced with respect to the signal of Eq.~(\ref{eq:sll}) by imposing
an additional cut $\theta_{\ell^+\ell^-}> \theta_{\ell^+\ell^-}^{\rm min}$~\cite{agashe}.
With an appropriate choice of $\theta_{\ell^+\ell^-}^{\rm min}$, the number of background events
$B_{\ell^+ \ell^-}^{(\ch^+\ch^-)}$ should be sufficiently reduced at least in the case of same
flavour leptons~\cite{agashe}.

A second source of background events comes from the neutralino decays into taus, followed by the 
decay of taus into leptons:
\begin{eqnarray}
B_{\ell^+ \ell^-}^{\tau\tau}
\!\!&=&\!\!
\sigma_{\nt_2}\times\epsilon_{\ell}^2\times\epsilon_{2\tau_\ell}\times\epsilon_{\rm cut}\times L\times
\nonumber \\
\!\!&\times&\!\!
{\rm BR}(\nt_2 \!\to\!\nt_1\tau\tau)\times\bigg[{\rm BR}(\tau\to\ell\nu\bar{\nu})\bigg]^2\!,
\label{eq:bg2}
\end{eqnarray}
where ${\rm BR}(\tau\to \ell \nu\bar{\nu})\simeq 0.17$ and $\epsilon_{2\tau_\ell}$ parameterizes
the acceptance of the two leptons from tau decays with respect to the acceptance of the two leptons directly produced in the decay chain of Eq.~(\ref{cascade}). Since the former leptons should be
softer than the latter, we expect $\epsilon_{2\tau_\ell}\lesssim 1$.
For the same reason, this source of background could be reduced by a cut on the invariant mass of
the two leptons. In our case, there is an additional source of background only for the $\mu^+\mu^-$ events: the flavour violating decay $\nt_2\to\nt_1\mu\tau$ followed by the decay of the tau
into a muon:
\begin{eqnarray}
B_{\mu^+ \mu^-}^{\tau\mu} &=&
\sigma_{\nt_2}\times\epsilon_{\ell}^2\times\epsilon_{1\tau_\ell}\times\epsilon_{\rm cut}\times L\times
\nonumber \\
&\times &{\rm BR}(\nt_2\!\to\!\nt_1\mu\tau)\,{\rm BR}(\tau\to\mu\nu\bar{\nu})\,,
\label{eq:bg3}
\end{eqnarray}
where $\epsilon_{1\tau_\ell}$ is the relative acceptance for a lepton coming from a tau decay.

Let us now consider the LFV neutralino decay. The total number of LFV $\tau^+\mu^-$ and $\tau^-\mu^+$ events is given by:
\begin{eqnarray}
S_{\tau\mu}&=&
2\times\sigma_{\nt_2}\times\epsilon_{\tau_h}\times\epsilon_{\ell}\times\epsilon_{\rm cut}\times L\times
\nonumber\\
&\times& {\rm BR}(\nt_2 \to \nt_1 \tau \mu)\,{\rm BR}(\tau\to h)\,,
\label{eq:staumu}
\end{eqnarray}
where only hadronically decaying taus $\tau\to h$ have been considered
(with ${\rm BR}(\tau\to h) \simeq 65\% $) and $\epsilon_{\tau_h}$ is the detection
efficiency for a tau jet.

Similarly to the flavour conserving case, the SUSY background is mainly given by
$\ch^+_1\ch^-_1$ decays and by di-tau neutralino events:
\begin{eqnarray}
B_{\tau\mu}^{\ch^{\!+}\!\ch^{\!-}\!} \!\!\!&=&\!\!
2\times\sigma_{\ch^+\ch^-}\times\epsilon_{\tau_h}\times\epsilon_{\ell}\times\epsilon_{\rm cut}
\times{\rm BR}(\tau\!\!\to\! h)\times  
\nonumber\\
\!\!\!&\times&\!\!\!
{\rm BR}(\ch^\pm_1\!\!\to\!\tau^\pm\! X)\,{\rm BR}(\ch^\pm_1\!\!\to\!\mu^\pm\! X)\,,
\label{eq:lfvbg1}
\end{eqnarray}
\begin{eqnarray}
B_{\tau\mu}^{\tau\tau} \!\!\!&=&\!\!
2\times\sigma_{\nt_2}\times\epsilon_{\ell}\times\epsilon_{\tau_h}\times\epsilon_{1\tau_\ell}
\times\epsilon_{\rm cut}\times L\times\nonumber \\
\!\!\!&\times&\!\!\!
{\rm BR}(\nt_2\!\to\!\nt_1\tau\tau)\,{\rm BR}(\tau\!\to\! h)\,
{\rm BR}(\tau\!\to\!\mu\nu\bar{\nu})\,.
\label{eq:lfvbg2}
\end{eqnarray}
As in the flavour conserving case, the background from $\ch^+\ch^-$ can be reduced with 
cuts on the angle between the muon and the tau-jet.

As we will see in the following, the number of background events of Eqs.~(\ref{eq:lfvbg1}, \ref{eq:lfvbg2}) can easily overwhelm the signal.
However, an equal number of $\tau-e$ events is expected to come from the processes contributing
to the $\tau-\mu$ background while, as previously mentioned, we cannot have simultaneously
sizeable LFV both in the $\tau-\mu$ and in the $\tau-e$ sectors because of the $\mu\to e\gamma$ constraints.
Therefore, the subtraction of the number of $\tau e$ events from the number of $\tau-\mu$
events should allow us to cancel the background, so that an excess of $\tau-\mu$ events
would be a signal of LFV \cite{hinchliffe,ellis}. In particular, the observation of $\tau-\mu$
LFV at the LHC should be possible, as long as ${\rm BR}(\nt_2\to\nt_1\tau\mu)/{\rm BR}(\nt_2\to\nt_1\tau\tau)\gtrsim 0.1$ \cite{ellis}.

\section{Representative points}\label{sec:6}
\begin{table*}[t]
\begin{center}
\begin{tabular}{|c|c|c|c|c|c|c|c|c|}
\hline
& $\sigma_{\rm SUSY}$ & $\sigma_{ee}$ & $\sigma_{\mu\mu}$ & $\sigma_{\tau\mu}$ & $|\Delta m_{\sle}/m_{\sle}|$ &
$|\Delta m_{ll}/m_{ll}|$ & $a^{\rm SUSY}_\mu$ & BR($\tau \to \mu \gamma$)  \\ 
\hline\hline
{\bf Point A} & ~5.2 pb~ & ~63 fb~&~ 43 fb ~& ~24 fb~ & ~1.1 \%~ & ~10 \%~ & ~ $1.2\times 10^{-9}$~ &~ $1.7\times 10^{-8}$~ \\
{\bf Point B} & ~1.8 pb~& ~32 fb~ & ~18 fb~& ~15 fb~ & ~1.3 \%~ & ~7.6 \%~ & ~ $8.0\times 10^{-10}$~ & ~$7.3\times 10^{-9}$~ \\
{\bf Point C} & ~9.7 pb~ & ~62 fb ~ & ~49 fb ~ & ~110 fb ~ & ~ 2.7 \%~ & ~ 4.9 \%~ & ~ $1.5\times 10^{-9}$~ & ~$2.4\times 10^{-8}$~ \\
{\bf Point D} & ~18.2 pb~ & ~169 fb~ & ~91 fb~ & ~536 fb~ & ~ 3.0 \%~ & ~ 6.2 \%~ & ~ $1.6\times 10^{-9}$~ & ~$1.3\times 10^{-8}$~\\
\hline
\end{tabular}
\end{center}
\caption{Predictions for low and high energy observables for the Points A,B,C,D discussed in the
         text.}
\label{tab:cross-sections}
\end{table*}
\begin{table*}[t]
\begin{center}
\begin{tabular}{|c||c|c|c|c||c|c|c||c|c|c|c|}
\hline
& $S_{\mu^+ \mu^-}$ & $B^{(\ch^+\ch^-)}_{\mu^+ \mu^-}$ & $B^{(\tau\tau)}_{\mu^+ \mu^-}$ &
$B^{(\tau\mu)}_{\mu^+ \mu^-}$& $S_{e^+ e^-}$ & $B^{(\ch^+\ch^-)}_{e^+ e^-}$ & $B^{(\tau\tau)}_{e^+ e^-}$ &
$S_{\tau\mu}$ & $B_{\tau\mu}^{(\ch^+\ch^-)}$ & $B_{\tau\mu}^{(\tau\tau)}$ & $\frac{{\rm BR}(\tau\mu)}{ {\rm BR}(\tau\tau)}$ \\
\hline\hline
{\bf Point A} & 850  & 0.65 $S_{\mu^+ \mu^-}$ & 0.12 $S_{\mu^+ \mu^-}$ & 0.09 $S_{\mu^+ \mu^-}$
& 1275   & 0.44 $S_{e^+ e^-}$ & 0.09 $S_{e^+ e^-}$
& 490   & 1.15 $S_{\tau\mu}$ & 1.3 $S_{\tau\mu}$ & 0.12 \\
{\bf Point B} & 364 & 0.64 $S_{\mu^+ \mu^-}$ & 0.07 $S_{\mu^+ \mu^-}$ & 0.14 $S_{\mu^+ \mu^-}$
& 648   & 0.35 $S_{e^+ e^-}$ & 0.04 $S_{e^+ e^-}$
& 307   & 0.82 $S_{\tau\mu}$ & 0.53 $S_{\tau\mu}$ & 0.32 \\
{\bf Point C} & 992 & 0.48 $S_{\mu^+ \mu^-}$ & 0.19 $S_{\mu^+ \mu^-}$ & 0.38 $S_{\mu^+ \mu^-}$
& 1255   & 0.38 $S_{e^+ e^-}$ & 0.15 $S_{e^+ e^-}$
& 1126   & 0.21 $S_{\tau\mu}$ & 0.5 $S_{\tau\mu}$ & 0.34 \\
{\bf Point D} & 1842 & 0.16 $S_{\mu^+ \mu^-}$ & 0.45 $S_{\mu^+ \mu^-}$ & 1.02 $S_{\mu^+ \mu^-}$
& 3822   & 0.09 $S_{e^+ e^-}$ & 0.24 $S_{e^+ e^-}$
& 10974   & 0.03 $S_{\tau\mu}$ & 0.44 $S_{\tau\mu}$ & 0.38 \\
\hline
\end{tabular}
\end{center}
\caption{Expected number of signal and background events for the relevant flavour conserving
and violating channels. The estimate has been done taking for the integrated luminosity
$L = 100~{\rm fb^{-1}}$.
}
\label{tab:s-b}
\end{table*}
We consider the following representative points of the parameter space studied in section \ref{Sec:4}
(for all $\tan\beta=10$, $A_0=0$):

{\bf Point A}: $m_0= 90$ GeV, $M_{1/2} = 400$ GeV and $(\delta_{\rm LL})_{32} = 0.03$.
This point lies in the neutralino-stau coannihilation region and reduces the
$(g-2)_\mu$ tension below the 2-$\sigma$ level. BR($\tau \to \mu \gamma$) is predicted
to be just below the present bound (see the top panel of Fig.~\ref{par-space}).

{\bf Point B}: $m_0= 105$ GeV, $M_{1/2} = 500$ GeV and $(\delta_{\rm LL})_{32} = 0.03$.
This point lies in the coannihilation region as shown by the top panel of
Fig.~\ref{par-space} but gives a smaller SUSY contribution to $(g-2)_\mu$ and
BR($\tau\to\mu\gamma$) compared to Point A.

{\bf Point C}: $m_0= 90$ GeV, $M_{1/2} = 350$ GeV and $(\delta_{\rm RR})_{32} = 0.1$.
This point lies in the neutralino-stau coannihilation region and reduces the $(g-2)_\mu$
tension below the 2-$\sigma$ level. BR($\tau\to\mu\gamma$) is predicted to be just below
the present bound (see the bottom panel of Fig.~\ref{par-space}).

{\bf Point D}: $m_0= 150$ GeV, $M_{1/2} = 300$ GeV, $(\delta_{\rm RR})_{32} = 0.1$.
This point does not lie in the coannihilation region, but it is still in the region
where the $\nt_2$ decays into ${\tilde e}_R$ and ${\tilde \mu}_R$ are kinematically allowed
(while the decays into LH sleptons are forbidden). The $(g-2)_\mu$ tension is reduced
below the 2-$\sigma$ level and BR($\tau\to\mu\gamma$) attains experimentally visible
values.

In Tab.~\ref{tab:cross-sections}, we present the relevant cross-sections of
Eqs.~(\ref{eq:cross-sections}) for the points listed above, as well as the corresponding
mass splitting $|\Delta m_{\sle}/m_{\sle}|$ and the edge splitting $|\Delta m_{ll}/m_{ll}|$.
The resulting $a^{\rm SUSY}_\mu$ and BR($\tau\to\mu\gamma$) are also provided.

Estimates of signals and backgrounds discussed in the previous section are given in Tab.~\ref{tab:s-b}. 
An integrate luminosity of $L = 100~{\rm fb^{-1}}$ has been assumed. 
We employed the results of the previous section together with numerical computation of SUSY production 
cross-section \cite{prospino} and decay branching fractions \cite{susyhit}. Following \cite{agashe}, 
we assumed the following values for the parameters: $\epsilon_{\rm cut}=1/4$, 
$\epsilon_{\ell} = 0.9$, $\epsilon_{2\tau_\ell} = \epsilon_{1\tau_\ell} =1$, $\epsilon_{\tau_h}=0.7$.

As a general comment, we see that the flavour conserving channels should be distinguished from the background,
especially if $B^{\ch^+\ch^-}_{\ell^+\ell^-}/S_{\ell^+\ell^-}$ could be reduced by a cut on the angle
between the leptons, $\theta_{\ell^+\ell^-}$ (clearly, this would reduce the number of the signal events, but 
increasing the S/B ratio). An exception is represented by point D: in this case the large background
$B^{(\tau\tau)}_{\mu^+ \mu^-}$ could make the measurement of the smuon mass very challenging, unless
a lower cut on the di-muon invariant mass could decrease the background. For the other points, it should be possible to extract the slepton masses from the di-lepton invariant mass distributions.
Of course, in order to do so, a quite large integrated luminosity (such as the 100 $\rm fb^{-1}$ considered in Tab.~\ref{tab:s-b}), i.e. some years of data taking, is probably necessary in order
to collect enough statistics.

Concerning the LFV channel, we see that the detection condition 
${\rm BR}(\nt_2 \to \nt_1 \tau \mu)/{\rm BR}(\nt_2 \to \nt_1 \tau \tau) \gtrsim 0.1$ \cite{ellis} is satisfied 
for all the points. In particular, detection of $\nt_2 \to \nt_1 \tau \mu$ looks very promising for Points C and D, 
due to the relatively low background and the large ratio 
${\rm BR}(\nt_2 \to \nt_1 \tau \mu)/{\rm BR}(\nt_2 \to \nt_1 \tau \tau)\simeq 0.3$. 

Let us now go in detail through the results presented in Tabs.~\ref{tab:cross-sections}-\ref{tab:s-b} in order to show, in
particular, the possible interplay among low-energy and collider searches for LFV. 

Let us first consider Point A. We see that the per-cent mass splitting between ${\tilde e}_L$ and ${\tilde \mu}_L$ induced by $(\delta_{\rm LL})_{32}$ results in a large 10 \% splitting of the
$e-e$ and $\mu-\mu$ edges. Moreover, we checked that an intermediate ${\tilde e}_R$ (${\tilde\mu}_R$) gives a contribution to $\Gamma(\nt_2 \to \nt_1 e^+e^-)$ ($\Gamma(\nt_2\to\nt_1 \mu^+\mu^-)$) which
is around 20 \%. As a consequence, the kinematical end points corresponding to ${\tilde e}_L$
and ${\tilde \mu}_L$ should not be hidden by a large number of events mediated by right-handed sleptons. Thus, we can expect $\Delta m_{\sle}/m_{\sle}$ to be measured at the LHC.
BR($\tau\to\mu\gamma$) is predicted
in the reach of the future experiments. Furthermore,
the LFV neutralino decay has also a good probability to be detected (we find
${\rm BR}(\nt_2\to\nt_1\tau^\pm\mu^\mp)\simeq 3 \%$, 
which corresponds to ${\rm BR}(\nt_2 \to \nt_1 \tau \mu)/{\rm BR}(\nt_2 \to \nt_1 \tau \tau)\simeq 0.1$ ). 
We can conclude that Point A provides an example
in which all the LFV observables we discussed so far could provide a positive signal.
Hence, in this case, the combined analysis of all the above observables would represent a
powerful tool in the attempt to reconstruct the flavour structure of the slepton sector.

Point B represents a case where the LHC could provide positive signals of LFV better than the
direct searches for $\tau\to\mu\gamma$, which is predicted to be beyond the reach of a
SuperB factory at KEK~\cite{KEK} but still within the reach of a Super Flavour Factory~\cite{super-flavour}. The measurement of the slepton mass splitting would for sure
require a larger integrated luminosity than what is required by Point A.
Still, the prospects at the LHC are reasonably favorable, given the low contribution of
the RH sleptons to the $\nt_2$ decay widths (at around the 10\% level) and the large LFV
rate ${\rm BR}(\nt_2 \to \nt_1 \tau^\pm\mu^\mp)\simeq 5 \%$.

Passing to Point C, we like to emphasize that, in spite of similar predictions for the production cross sections to the Point A, we find that the measurement of the mass splitting between
${\tilde e}_R$ and ${\tilde \mu}_R$ is more challenging. The reason is that we are in a kinematic region where the decays of $\nt_2$ into LH sleptons are open. As a consequence, since $\nt_2$ is mostly Wino, the contribution to the decay widths $\Gamma(\nt_2 \to \nt_1 e^+e^-)$ and $\Gamma(\nt_2 \to \nt_1 \mu^+\mu^-)$ from intermediate LH sleptons is almost twice the contribution from
${\tilde e}_R$ and ${\tilde \mu}_R$.
Even though the production cross sections that are effective for the measurement of the mass
splitting for right-handed sleptons are much smaller than the total production cross sections
(that are dominated by the contributions of LH sleptons), we notice that the kinematical end
points for the invariant mass distributions of right-handed sleptons appear at higher values
compared to those of left-handed sleptons; as a result, the measurements of the kinematic
edges associated with the RH sleptons should not be affected by a large background.
Therefore, $\Delta m_{\sle}/m_{\sle}$ could be still measured provided enough statistics.
Moreover, Point C gives a very large rate to the LFV $\nt_2$ decay (being ${\rm BR}(\nt_2\to\nt_1 \tau^\pm\mu^\mp)\simeq 7 \%$) and a prediction for BR($\tau\to\mu\gamma$) in the reach of the
future experiments.

Finally, Point D, being in a region where the $\nt_2$ decays into LH sleptons are not allowed,
could give good prospects for the measurement of the ${\tilde e}_R$-${\tilde \mu}_R$ mass splitting,
only if the background to the $\mu\mu$ events coming from the LFV neutralino decay 
($B^{(\tau\mu)}_{\mu^+ \mu^-}$) could reduced with a lower cut on the di-muon invariant mass.
Another caveat is that around the 30 \% of the ``pure'' $\mu-\mu$ 
events are mediated by $\tilde{\tau}_1$, so that the di-muon distribution
will exhibit two edges, one corresponding to the kinematical end-point of ${\tilde \mu}_R$, the
other one corresponding to $\tilde{\tau}_1$.
On the other hand, the huge cross-section of $\nt_2 \to \nt_1 \tau^\pm\mu^\mp$ (up to a factor of
3-6 larger than the cross-sections of the flavour conserving decays) and the relatively low background make the LFV $\nt_2$ decay the most sensitive LFV observable.

\section{Conclusions}\label{Sec:7}

In this paper, we have studied the sensitivity of the slepton mass splittings
of the first two generations to LFV effects.

In particular, we have considered minimal SUGRA scenarios, predicting highly degenerate
first two slepton generations. Hence, any experimental evidence for a selectron/smuon
mass splitting $\Delta m_{\sle}/m_{\sle}$, say above the percent level, would signal
either a different mechanism for the SUSY breaking or non minimal realizations of SUGRA 
models. In the latter case, the presence of LFV interactions might explain the 
potential evidence of a significant $\Delta m_{\sle}/m_{\sle}$, as throughly discussed 
in this paper.

Interestingly enough, we have found that sizable values for $\Delta m_{\sle}/m_{\sle}$
can be generated only through flavour mixings between the second and third slepton
families, a scenario that naturally arises and is well motivated by the large mixing
angle observed in atmospheric neutrino oscillation experiments.

In contrast, flavour mixings between the second and first slepton families are highly
constrained by the limit on $\rm{BR}(\mu\to e\gamma)$ and they can hardly account for
such a mass splitting \cite{Hisano:2002iy}.

We have shown that slepton mass splittings may be competitive with low energy processes
to probe lepton flavour violation in large regions of the SUSY parameter space.


 Moreover, we have exploited the NP sensitivity of the high energy LFV process
$\nt_{2}\to\nt_{1}\tau^{\pm}\mu^{\mp}$ assuming an integrated luminosity $L=100~{\rm fb^{-1}}$.
We have found that, whenever the relevant background is kept under control by means of
appropriate kinematical cuts, $\nt_{2}\to\nt_{1}\tau^{\pm}\mu^{\mp}$ can be experimentally
visible in sizable regions of the parameter space where, at the same time:

(i) BR$(\tau\to\mu\gamma)$ can be below the current experimental bound and even 
beyond the reach of a SuperB factory at KEK (BR$(\tau\to\mu\gamma)>10^{-8}$);

(ii) $a^{\rm SUSY}_{\mu}\gtrsim 1\times 10^{-9}$ providing an explanation for the 
$(g-2)_{\mu}$ anomaly, at least at the 2-$\sigma$ level;

(iii) the WMAP relic density constraint can be fulfilled by an effective neutralino-stau
coannihilation;

(iv) the LFV induced mass splitting between selectron and smuons can be measured at the LHC.

Hence, $\nt_{2}\to\nt_{1}\ell_i\ell_j$ and $\tau\to\mu\gamma$ might be complementary
probes of LFV in SUSY with $\nt_{2}\to\nt_{1}\tau^{\pm}\mu^{\mp}$ being even more powerful
than $\tau\to\mu\gamma$ for a heavy SUSY spectrum.

In conclusion, the present study represents another proof of the synergy and interplay
existing between the LHC, i.e. the {\em high-energy frontier}, and high-precision
low-energy experiments, i.e. the {\em high-intensity frontier}.\\

%
\textit{Acknowledgments:}
We thank M.~Nagai for useful discussions.
This work has been supported in part by the Cluster of Excellence ``Origin and Structure
of the Universe'' and by the German Bundesministerium f{\"u}r Bildung und Forschung under
contract 05H09WOE.



\end{document}